\documentclass[astrosymb, longbib, tighten, twocolappendix, twocolumn]{aastex701}

\usepackage[version=4]{mhchem}
\usepackage{csquotes}
\usepackage{siunitx}

\DeclareSIUnit[number-unit-product=\,]{\molecule}{molecule}

\received{June 3, 2026}

\submitjournal{The Astrophysical Journal}

\begin{document}

\title{Ultraviolet-Driven Atmospheric Degeneracies Challenge Conventional Biosignature Frameworks for Terrestrial Planets with Ultracool M Dwarf Hosts: An Archean-Analog TRAPPIST-1 e Case Study}

\author[0000-0001-5290-1001]{Evan L. Sneed}
\affiliation{Department of Earth and Planetary Sciences, University of California, Riverside, CA 92521, USA}
\email[show]{evan.sneed@email.ucr.edu}

\correspondingauthor{Evan L. Sneed}

\author[0000-0002-2949-2163]{Edward W. Schwieterman}
\affiliation{Department of Earth and Planetary Sciences, University of California, Riverside, CA 92521, USA}
\affiliation{Blue Marble Space Institute of Science, Seattle, WA 98104, USA}
\email[show]{eschwiet@ucr.edu}

\author[0000-0002-1046-025X]{Sarah R. Peacock}
\affiliation{University of Maryland, Baltimore County, MD 21250, USA}
\affiliation{NASA Goddard Space Flight Center, Greenbelt, MD 20771, USA}
\email{speacock@umbc.edu}

\author[0000-0002-0413-3308]{Nicholas F. Wogan}
\affiliation{SETI Institute, Mountain View, CA 94043, USA}
\affiliation{NASA Ames Research Center, Moffett Field, CA 94035, USA}
\email{nicholas.f.wogan@nasa.gov}

\author[0000-0001-8674-6775]{Timothy W. Lyons}
\affiliation{Department of Earth and Planetary Sciences, University of California, Riverside, CA 92521, USA}
\email{timothyl@ucr.edu}

\begin{abstract} 

The ultraviolet (UV) spectrum of a host star strongly shapes the atmospheric composition and potential biosignatures of its planets. This relationship may be especially important for the planets orbiting TRAPPIST-1, an M8V star with substantially different published UV spectral energy distributions (SEDs). Using a one-dimensional photochemical model, we quantify how these SED uncertainties affect Archean Earth–like atmospheric analogs on TRAPPIST-1 e with and without biospheres. We emphasize Earth’s Archean epoch because it represents a planet in transition from primarily abiotic to biotic controls on atmospheric composition. Different stellar spectra produce order-of-magnitude variations in the predicted abundances of \ce{CH4}, \ce{CO}, \ce{O2}, and \ce{O3}, thereby generating photochemical degeneracies that complicate the interpretation of potential biosignatures. For one TRAPPIST-1 UV reconstruction, a modeled atmosphere with abiotic deposition velocities and low \ce{CH4} input can sustain simultaneous spectrally discernible \ce{CH4} and \ce{O3}, yielding a potential false-positive disequilibrium biosignature. For all SEDs tested, surface deposition consistent with microbially-mediated \ce{CO} consumption allows substantial \ce{O2} and \ce{O3} accumulation even without oxygenic photosynthesis, implying that oxygen-rich atmospheres around ultracool M dwarfs may not uniquely trace oxygenic ecosystems. Across our models, \ce{CO} remains a powerful discriminator between abiotic and biotic surface boundary assumptions. Overall, we show that the abundances of co-occurring \ce{CH4}, \ce{CO}, and \ce{O3} can vary by orders of magnitude, depending on the assumed stellar spectrum, creating ambiguities in interpreting atmospheric biosignatures, though observability may be challenging with current capabilities. Reducing UV spectral uncertainties is therefore essential for assessing surface-to-atmosphere interactions of temperate exoplanets around ultracool M dwarfs.  

\end{abstract}


\makeatletter\def\Hy@Warning#1{}\makeatother

\section{Introduction}

The TRAPPIST-1 system (d $\approx$ 12.5 pc) remains a major benchmark for assessing the habitability and atmospheric evolution of terrestrial planets around ultracool M dwarfs \citep{gillonTemperateEarthsizedPlanets2016, gillonSevenTemperateTerrestrial2017}. With four planets (d–g) in or near the habitable zone \citep{ducrotTRAPPIST1GlobalResults2020, agolRefiningTransittimingPhotometric2021}, it has become a focal point for astrobiology. Yet the same stellar environment that makes TRAPPIST-1 so valuable may also be hostile to long-lived surface habitability: frequent energetic flares and sustained XUV (X-ray and extreme-UV) emission can enhance atmospheric escape, drive water loss, and promote atmospheric and surface oxidation on close-in planets \citep{vidaFrequentFlaringTRAPPIST12017, venotINFLUENCEStelLARFLARES2016, doamaralContributionMdwarfFlares2022, krissansen-tottonImplicationsAtmosphericNondetections2023, zahnleCosmicShorelineEvidence2017, passRecedingCosmicShoreline2025}. The compact architecture of the system, with orbital periods of $\sim$1.5–19 days, also suggests that TRAPPIST-1's planets may rotate synchronously, concentrating stellar irradiation on the day side of the planet and further influencing the lifetime and circulation of atmospheric gases \citep{agolRefiningTransittimingPhotometric2021}. These secondary atmospheres could persist if volatile inventories are sufficiently large and if atmospheric circulation can efficiently redistribute heat across the entire planetary surface \citep{wordsworthATMOSPHERICHEATREDISTRIBUTION2015,auclair-desrotourAtmosphericStabilityCollapse2020, thomasStatisticalGeochemicalConstraints2025}. Increasingly precise transit and eclipse measurements are now beginning to probe whether these theoretical expectations match our observations of these atmospheres.

\begin{figure*}[t!] 
    \centering
    \includegraphics[width=\linewidth]{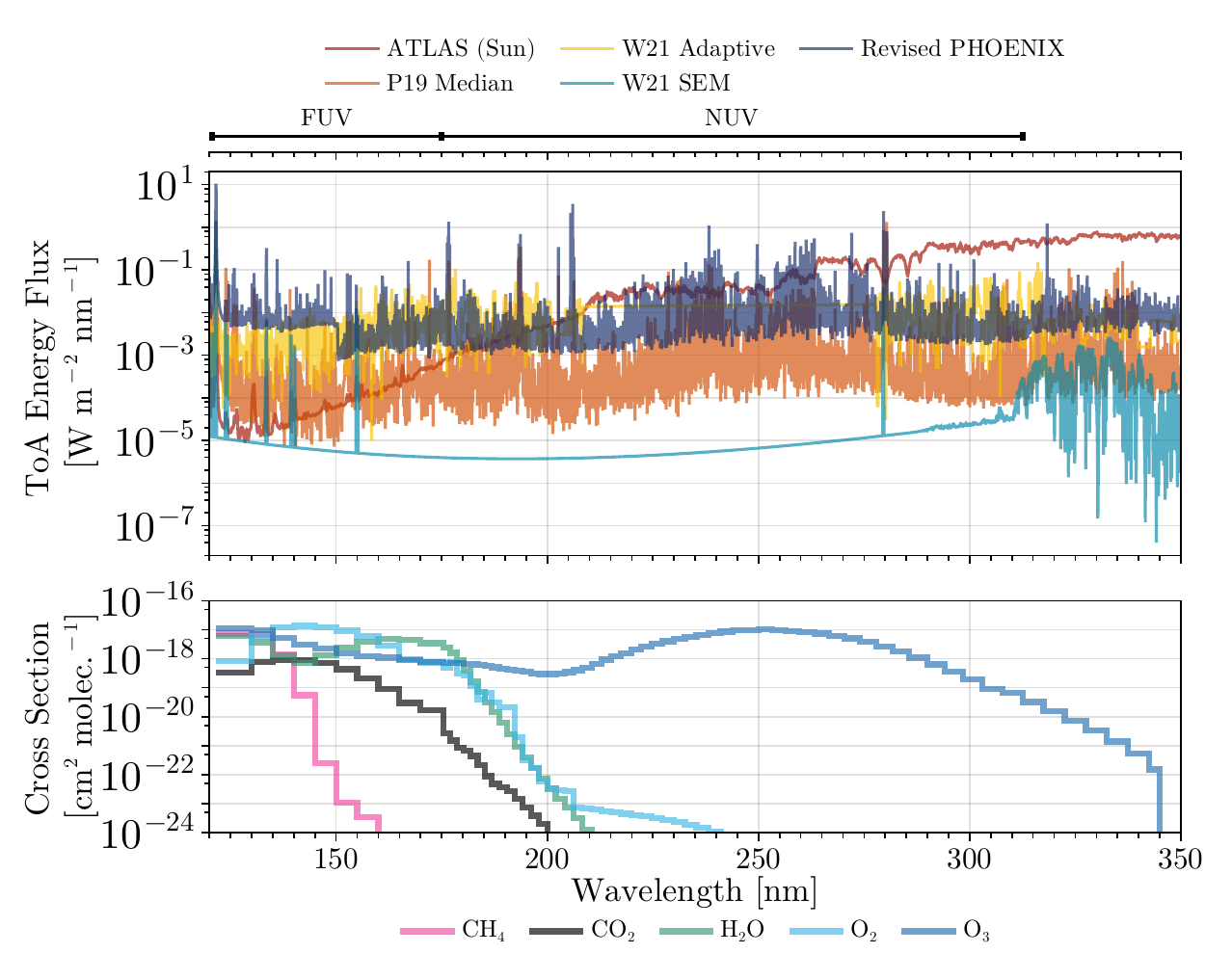}
    \caption{Comparison of input SEDs used in this study and molecular photoabsorption cross sections relevant to Archean-analog photochemistry. The horizontal bars indicate the FUV and NUV wavelength intervals used to compute the stellar FUV/NUV ratios in Figures~\ref{fig:fuvnuv} and \ref{fig:fuvnuv_column_densities}. \textbf{Top:} Top-of-atmosphere (ToA) stellar energy fluxes at TRAPPIST-1 e from 120–350 nm for five SEDs: Peacock+2019 (P19) Median (orange) \citep{peacockPredictingExtremeUltraviolet2019}, Mega-MUSCLES (W21) Adaptive (yellow), Mega-MUSCLES (W21) SEM (light blue) \citep{wilsonMegaMUSCLESSpectralEnergy2021}, a \enquote{Revised PHOENIX} model (dark blue) constrained by the HST observations published in \citet{wilsonMegaMUSCLESSpectralEnergy2021}, and a scaled modern Sun (red) \citep{thuillierSolarIrradianceReference2004}. All SEDs are scaled to TRAPPIST-1 e's ToA irradiance. \textbf{Bottom:} Representative molecular cross sections for \ce{CH4} (pink), \ce{CO2} (black), \ce{H2O} (green), \ce{O2} (light blue), and \ce{O3} (dark blue) are shown, highlighting the wavelength overlap between the cross sections and the stellar UV output. Where both the flux and the cross section are large, photolysis rates increase and influence production and loss pathways throughout the atmosphere.}
    \label{fig:stellar_spectra}
\end{figure*}

Observationally, a coherent picture is emerging for the inner planets. Pre-JWST transmission spectra for TRAPPIST-1 b, c, and d were featureless at low resolution, excluding cloud-free \ce{H2}-rich envelopes \citep{dewitCombinedTransmissionSpectrum2016, dewitAtmosphericReconnaissanceHabitablezone2018}. JWST thermal-phase and eclipse measurements now indicate that b is most consistent with an airless rocky surface, with thick atmospheres strongly disfavored and only a narrow range of tenuous, low-opacity alternatives not yet fully excluded \citep{greeneThermalEmissionEarthsized2023, gillonNoThickAtmosphere2026}. For c, new phase-curve constraints disfavor both a thick \ce{CO2}-dominated atmosphere and most other dense-atmosphere scenarios, while still permitting either an airless, relatively reflective surface or a low-opacity secondary atmosphere, such as a\ce{O2}-dominated or low-pressure atmosphere containing only small amounts of \ce{CO2} and \ce{H2O} \citep{ziebaNoThickCarbon2023, gillonNoThickAtmosphere2026}. For d, corrected transmission spectra show no molecular features, tightening limits on high-mean-molecular-weight atmospheres \citep{piaulet-ghorayebStrictLimitsPotential2025}. For potentially habitable TRAPPIST-1 e, recent measurements rule out \ce{H2}-rich envelopes and disfavor \ce{CO2}-dominated cases, though \ce{N2}-dominated secondary atmospheres with mixtures of \ce{CH4} and trace \ce{CO2} remain viable when spectral interpretation attempts to control for stellar contamination \citep{gliddenJWSTTSTDREAMSSecondary2025, espinozaJWSTTSTDREAMSNIRSpec2025}.

Photochemical studies of TRAPPIST-1's habitable-zone planets have explored a broad range of atmospheric scenarios, including desiccated \ce{O2}-rich and Venus-like cases, as well as \ce{CO2}-\ce{N2}, \ce{CO2}-\ce{O2}, prebiotic, and modern Earth-like compositions \citep{omalley-jamesUVSurfaceHabitability2017, lincowskiEvolvedClimatesObservational2018, wunderlichDistinguishingWetDry2020, linDifferentiatingModernPrebiotic2021a}. Another relevant comparison case is the Archean Earth, an anoxic planetary state spanning from 4.0 to 2.5 Ga. This interval contains some of the earliest, most widely accepted evidence for life \citep{allwoodStromatoliteReefEarly2006, noffkeMicrobiallyInducedSedimentary2013, cavalazziCellularRemains342billionyearold2021, wolfeHorizontalGeneTransfer2018, wongOrganicGeochemicalEvidence2025}. Compared to the modern Earth's atmosphere, the Archean atmosphere likely featured higher \ce{CO2}, robust \ce{CH4} from microbial methanogenesis, negligible \ce{O2} and \ce{O3}, and comparable or lower \ce{N2} concentrations \citep{kharechaCoupledAtmosphereEcosystem2005, catlingArcheanAtmosphere2020, chenNewEmpiricalKinetics2026}. The Archean Earth likely represents a world in transition from one with primarily abiotic controls on atmospheric composition to one where life began to play a prominent role \citep{ jonesEvolutionEarthsAtmosphere2025}. Archean-like atmospheric compositions provide a physically motivated counterpoint to modern-Earth-like assumptions for temperate TRAPPIST-1 worlds and motivate the absence of an imposed biological \ce{O2} surface flux in this work \citep{mak3DSimulationsTRAPPIST1e2024, eager-nashSimulatingBiosignaturesPreoxygen2024}.

A central uncertainty in such modeling is the star’s UV spectral energy distribution (SED), which sets key photolysis rates and atmospheric lifetimes \citep{franceMUSCLESTREASURYSURVEY2016,kozakisOzoneReliableProxy2022}. Several distinct TRAPPIST-1 SEDs are now used in the literature. 
\cite{peacockPredictingExtremeUltraviolet2019} computed a series of three \texttt{PHOENIX} atmosphere models of TRAPPIST-1 calibrated to either the \cite{bourrierReconnaissanceTRAPPIST1Exoplanet2017} Ly$\alpha$ flux of TRAPPIST-1 or distance-adjusted GALEX photometry of similar M8 stars. As part of the Mega-MUSCLES (Measurements of the Ultraviolet Spectral Characteristics of Low-mass Exoplanetary Systems) Treasury survey, \cite{wilsonMegaMUSCLESSpectralEnergy2021} obtained HST Cosmic Origins Spectrograph (COS) and Space Telescope Imaging Spectrograph (STIS) measurements of TRAPPIST-1, resolving several FUV emission lines and the Mg II h \& k doublet in the NUV. With these data, they released two alternative SEDs: the Mega-MUSCLES Adaptive panchromatic SED, intended to represent quiescent output, and the Mega-MUSCLES Semi-empirical Model (SEM) variant, which replaces low-S/N \qtyrange{110}{420}{\nano\meter} data with a smoothed polynomial, reconstructed Ly$\alpha$, and synthetic emission lines.
\citet{wilsonMegaMUSCLESSpectralEnergy2021} recommended the SEM SED for planetary atmosphere studies because it replaces the noisiest observed UV regions with a semi-empirical reconstruction. However, the SEDs differ substantially across the far- and near-UV, where the photoabsorption cross sections of \ce{CH4}, \ce{CO2}, \ce{H2O}, \ce{O2}, and \ce{O3} are large. In particular, the Adaptive and SEM SEDs differ in the relative distribution of flux between the FUV (\qtyrange{120.0}{175.0}{\nano\meter}) and NUV (\qtyrange{175.0}{312.5}{\nano\meter}), a distinction that can strongly affect oxidative photochemistry \citep{harmanAbioticO2Levels2015}.

Prior studies already indicate that this sensitivity matters. One-dimensional photochemical models have shown that species such as \ce{O3}, \ce{H2O}, and hydrocarbon hazes respond strongly to UV flux differences between stars \citep{seguraBiosignaturesEarthLikePlanets2005, grenfellSensitivityBiosignaturesEarthlike2014, tealEffectsUVStellar2022}. Recent TRAPPIST-1 e modeling has further shown that pre-oxygenic biospheres can permit surface \ce{O2} accumulation alongside substantial atmospheric \ce{CO}, reinforcing the need to interpret Archean-analog biosignatures in the joint context of stellar UV forcing and lower boundary conditions \citep{eager-nashSimulatingBiosignaturesPreoxygen2024}. Additionally, three-dimensional \texttt{WACCM6} simulations of modern- and Proterozoic-Earth (representing the epoch from 2.5 to 0.54 Ga) analogs with oxygenated atmospheres around TRAPPIST-1 found that different stellar SEDs yield divergent photochemical predictions, particularly for atmospheric \ce{O3} \citep{cookeDegenerateInterpretationsO32023}. However, to our knowledge, existing terrestrial three-dimensional photochemical-climate models cannot yet self-consistently simulate Archean-like atmospheres without prescribed \ce{O2} levels and thus cannot determine how both \ce{O2} and \ce{O3} respond to stellar SED choice in otherwise anoxic \ce{N2}-\ce{CO2}-\ce{H2O} atmospheres with and without life.

In this study, we quantify how plausible TRAPPIST-1 UV SED choices alone drive order-of-magnitude differences in predicted atmospheric photochemistry on an Archean-analog TRAPPIST-1 e with and without life. Using a one-dimensional photochemical model, we compare steady-state abundances of \ce{CH4}, \ce{CO}, \ce{O2}, and \ce{O3} and compute transmission spectra across multiple input SEDs. We show that high-FUV/low-NUV cases can produce detectable \ce{O3} without life while simultaneously enhancing atmospheric \ce{CO}, complicating the common \ce{CH4}+\ce{O3} biosignature heuristic and creating false-positive and false-negative degeneracies for temperate planets orbiting ultracool M dwarfs. We then identify observational discriminants, particularly joint constraints on \ce{CO}, \ce{CH4}, and \ce{O3} spanning the JWST NIRSpec and MIRI wavelength ranges. Finally, we demonstrate that in inhabited worlds, substantial \ce{O2} and \ce{O3} concentrations can arise for modeled cases with low \ce{CH4} fluxes, without oxygenic photosynthesis, through the adoption of surface deposition consistent with microbial \ce{CO} consumption.

\section{Methods} \label{sec:methods}
 
\subsection{Stellar Spectra Modeling, Processing, and Binning} \label{ssec:stellar_spectra}

We explore the impact of using five distinct SEDs as stellar input in our one-dimensional photochemical model (Figure~\ref{fig:stellar_spectra}): a median spectrum of the three \citealt{peacockPredictingExtremeUltraviolet2019} \texttt{PHOENIX} models, both the Mega-MUSCLES Adaptive and SEM SEDs from \citealt{wilsonMegaMUSCLESSpectralEnergy2021}, a \enquote{Revised PHOENIX} model described below, and a scaled modern solar reference spectrum based on high-resolution ATmospheric Laboratory for Applications and Science (ATLAS-1) composite solar irradiance data \citep{thuillierSolarIrradianceReference2004}.

The \enquote{Revised PHOENIX} spectrum is an update motivated by the Mega-MUSCLES observations obtained after the initial TRAPPIST-1 \texttt{PHOENIX} models of \cite{peacockPredictingExtremeUltraviolet2019}. The more recent HST data provide improved empirical constraints on the UV spectrum, particularly in the FUV and for the Mg II h \& k doublet in the NUV, necessitating an updated model. However, this version has not yet been fully validated, and we therefore treat it as an illustrative end-member rather than a definitive reconstruction of TRAPPIST-1's UV output.

We construct a revised \texttt{PHOENIX} atmosphere model for TRAPPIST-1 using stellar parameters from \cite{agolRefiningTransittimingPhotometric2021} ($T_{\rm eff}=2566$ K, $\log g=5.24$), with HST/COS and STIS UV observations providing empirical constraints on the chromosphere and transition region. The model follows the general framework of \cite{peacockPredictingExtremeUltraviolet2019}, consisting of a photosphere with parametrized temperature rises in the chromosphere and transition region defined in column mass space. The transition region is extended to $2\times10^{5}$ K to reproduce high-temperature emission lines observed in the UV. Relative to the \cite{peacockPredictingExtremeUltraviolet2019} models, we include an expanded non-LTE treatment (78 versus 63 ionization stages) and updated atomic processes, including additional collisional rates with hydrogen and improvements to the treatment of the \ion{H}{1} Lyman-$\alpha$ line core. 

To identify a representative spectrum, we computed a grid of 72 models spanning a range of chromospheric and transition region thicknesses and locations. The best-fit model was selected via $\chi^2$ minimization against a set of well-resolved UV emission lines in the HST data, using continuum-normalized line fluxes measured over consistent wavelength intervals. The resulting model reproduces the FUV pseudocontinuum and matches 7 of 13 diagnostic line fluxes within observational uncertainties, with most remaining lines within a factor of a few. 

In terms of band-integrated UV fluxes, the \enquote{Revised PHOENIX} spectrum is broadly consistent with the Mega-MUSCLES Adaptive SED in the FUV and NUV, but with subtle systematic differences. The FUV flux is in close agreement with the empirical reconstruction, differing at the $\sim$0.1 dex level. In the NUV, the agreement is similarly good (within $\sim$0.1 dex), although constraints are weaker due to the limited signal-to-noise of the HST/STIS observations and the lack of strong continuum diagnostics beyond the Mg II h \& k doublet.

Figure~\ref{fig:stellar_spectra} shows the scaled SEDs, illustrating the UV structure relevant to the photochemical calculations. For input into \texttt{Atmos}, the TRAPPIST-1 SEDs were subsequently re-binned to the model wavelength grid with flux-conserving resampling using \texttt{SpectRes}. Spurious negative-flux bins were removed within the Mega-MUSCLES Adaptive data before rebinning was completed. Across UV wavelengths, the SEDs vary significantly, with FUV/NUV flux ratios spanning two orders of magnitude across the TRAPPIST-1 SEDs. The spectra also differ in the relative strengths of individual emission features. In the rebinned spectra, the Ly~$\alpha$ line at \qty{121.6}{\nano\meter} ranges from being only modestly enhanced above the neighboring FUV continuum in the Mega-MUSCLES Adaptive SED to being enhanced by roughly two orders of magnitude in the Mega-MUSCLES SEM SED. Similarly, the Mg II h \& k region near \qty{280}{\nano\meter} is prominent in several of the SEDs, but its contrast relative to the adjacent NUV continuum varies substantially across the TRAPPIST-1 SEDs.  

The photochemical importance of these spectral differences depends strongly on wavelength. In  Figure~\ref{fig:stellar_spectra}, we also show the absorption cross sections of key atmospheric species to highlight where differences among the adopted SEDs overlap with the UV bands responsible for photolysis of \ce{CH4}, \ce{CO2}, \ce{H2O}, \ce{O2}, and \ce{O3}. Between \qtyrange{120}{160}{\nano\meter}, the stellar flux overlaps with large photoabsorption cross sections for \ce{CH4}, \ce{CO2}, \ce{H2O}, and \ce{O2}; differences in this interval therefore affect direct \ce{CH4} destruction, \ce{CO2} photolysis to \ce{CO} and \ce{O}, and the production of H-bearing radicals from \ce{H2O}.
From roughly \qtyrange{160}{210}{\nano\meter}, \ce{CH4} absorption decreases rapidly, but \ce{CO2}, \ce{H2O}, and \ce{O2} absorption remain relevant, so this interval continues to influence the balance between oxidant production and \ce{HO_x}-mediated recycling of \ce{CO} and \ce{CH4}.
At longer wavelengths, especially from \qtyrange{200}{350}{\nano\meter}, \ce{O3} is the dominant absorber among the species shown, making the NUV continuum and strong features such as Mg II h \& k important for \ce{O3} photolysis.
Thus, spectra such as Mega-MUSCLES SEM, which retain short-wavelength FUV emission while suppressing much of the longer-wavelength UV continuum, can favor \ce{CO2}-driven oxidant production while reducing the UV pathways that would otherwise enhance radical production and \ce{O3} destruction.

\subsection{Photochemical Modeling with \texttt{Atmos}} \label{ssec:photochemical_methods}

\begin{figure*}[t!]
    \centering
    \includegraphics[width=\linewidth]{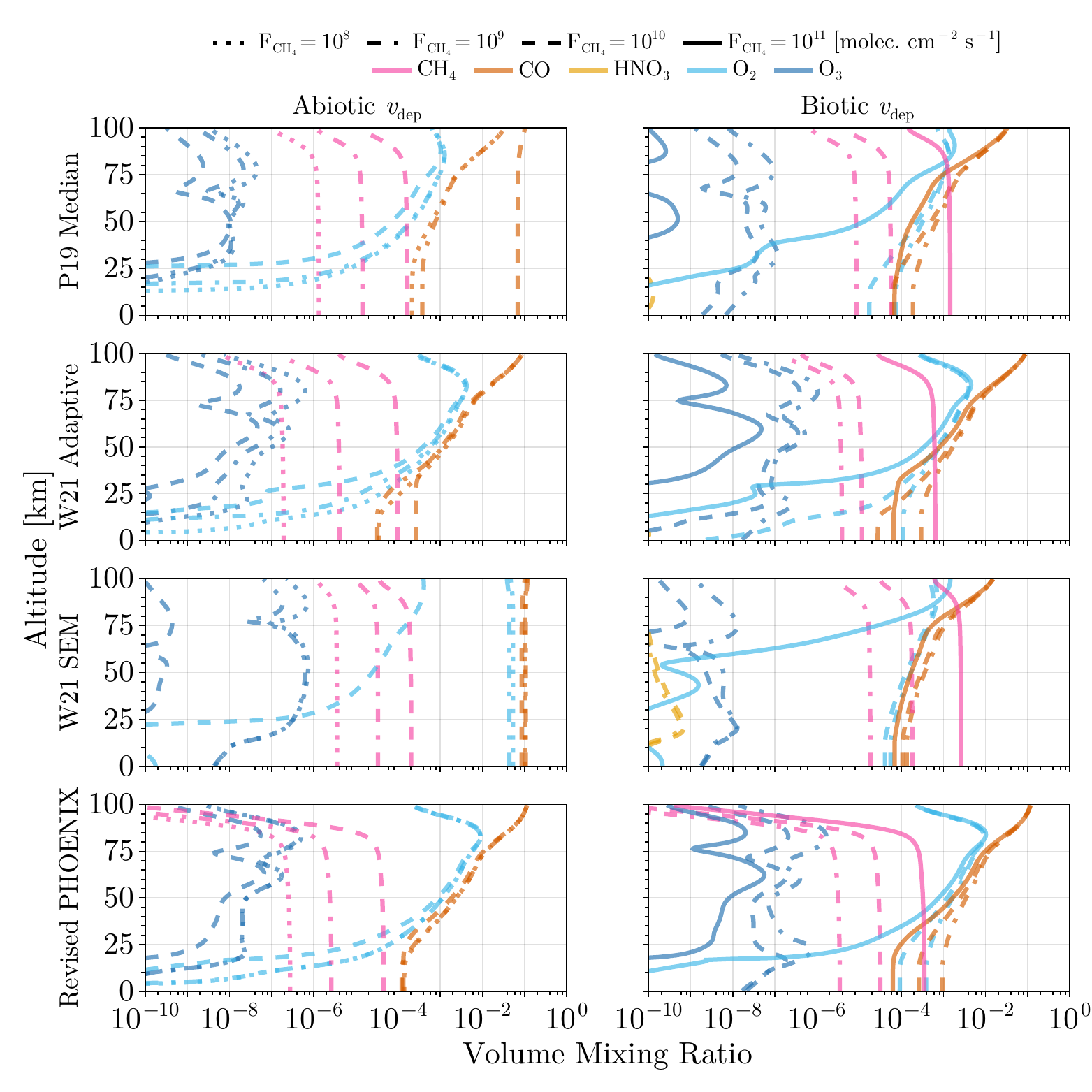}
    \caption{Vertical volume mixing ratio profiles of \ce{CH4}
    (pink), \ce{CO} (orange), \ce{HNO3} (yellow), \ce{O2} (light blue), and \ce{O3} (dark blue), from
    \texttt{Atmos} photochemical simulations. Line styles denote imposed methane surface fluxes $F_{\ce{CH4}}=10^{8}$ (dotted), $F_{\ce{CH4}}=10^{9}$ (dash-dotted), $10^{10}$ (dashed), and $10^{11}$ 
    molecules cm$^{-2}$ s$^{-1}$ (solid). The rows show four TRAPPIST-1 spectral energy
    distributions (top to bottom: Peacock+2019 (P19) Median, Mega-MUSCLES (W21) Adaptive, Mega-MUSCLES (W21) SEM, and
    \enquote{Revised PHOENIX}) while the columns compare abiotic (left) and biotic (right) deposition velocities. \ce{CH4} fluxes of $10^{8}$ through $10^{10}$ molecules cm$^{-2}$ s$^{-1}$ are shown in the abiotic column, while fluxes of $10^{9}$ through $10^{11}$ molecules cm$^{-2}$ s$^{-1}$ are shown in the biotic column. For the biotic Mega-MUSCLES Adaptive case, the nominal $F_{\ce{CH4}}=10^{10}$ molecules cm$^{-2}$ s$^{-1}$ run did not converge in \texttt{Atmos}; we therefore show the nearest converged solution, $F_{\ce{CH4}}=1.5\times10^{10}$ molecules cm$^{-2}$ s$^{-1}$. Higher $
    F_{\ce{CH4}}$ increases atmospheric \ce{CH4} and can drive
    \ce{CO} buildup in cases with abiotic deposition velocities, whereas enhanced deposition of \ce{CO} in the cases with biotic deposition velocities suppresses \ce{CO} buildup. \ce{O2} and \ce{O3} mixing ratios are generally higher at lower $F_{\ce{CH4}}$ and for spectra with elevated FUV/NUV ratios.}
    \label{fig:spaghetti}
\end{figure*}

We simulate Archean-analog atmospheres for TRAPPIST-1 e with the publicly available one-dimensional photochemical model included within \texttt{Atmos}. The model descends from the Kasting group's one-dimensional photochemical framework \citep{kastingOxygenLevelsPrebiological1979, pavlovUVShieldingNH32001, zahnleLossMassindependentFractionation2006}. Subsequent updates to the model's kinetics, radiative transfer, and wavelength-dependent cross sections have since been made. \texttt{Atmos} has been extensively used within the astrobiology community, including for simulations of the atmosphere of the Archean Earth and of Archean-analog exoplanet atmospheres \citep{arneyPaleOrangeDot2016, arneyPaleOrangeDots2017, tealEffectsUVStellar2022, wolfChemistryClimateTransmission2025}. The model integrates the coupled continuity equations for all atmospheric species to steady state, balancing chemical and photolytic production and loss with vertical transport by eddy and molecular diffusion. Photolysis rates are computed from wavelength-dependent absorption cross sections and quantum yields convolved with the input stellar spectral energy distributions, which directly link the stellar UV spectrum to the atmospheric composition.

All modeled atmospheres assume a surface pressure of 1~bar with a partial pressure of \qty{0.1}{bar} \ce{CO2}, a planetary radius fixed to the measured value of TRAPPIST-1~e, and a surface gravity consistent with its measured mass and radius \citep{agolRefiningTransittimingPhotometric2021}.
Following the methods of \citet{kopparapuHabitableZonesMainSequence2013}, we set the surface temperature to \qty{288.4}{\kelvin} and adopt a temperature–pressure profile consisting of a moist pseudoadiabat from the surface up to an isothermal \qty{180}{\kelvin} stratosphere. The adopted atmospheric fluxes, surface mixing ratios, and dry deposition velocities used in the model are provided in Table~\ref{tab:conditions}.

Alongside the previously available modern solar reference spectrum based on the ATLAS-1 composite solar irradiance spectrum of \citet{thuillierSolarIrradianceReference2004} and the Peacock+2019 Median SED, we import the three generated TRAPPIST-1 SEDs (Mega-MUSCLES Adaptive, Mega-MUSCLES SEM, and \enquote{Revised PHOENIX}) into \texttt{Atmos} to determine the extent of the atmospheric photochemical differences as a function of the differences in the UV region of these spectral energy distributions, especially as the UV helps to drive the core \ce{CO2}-\ce{CH4} photochemical-redox network. In the relevant FUV bands, photolysis of \ce{CO2} produces \ce{CO} and atomic oxygen. Atomic oxygen recombines to form \ce{O2}, and \ce{O2} participates in Chapman chemistry to form \ce{O3} \citep{harmanAbioticO2Levels2015}. The direct reaction of \ce{CO} with \ce{O} is spin-forbidden and inefficient, so \ce{CO} tends to accumulate in the cases with abiotic deposition velocities unless catalytic cycles or surface sinks return it to \ce{CO2}. The dominant gas-phase removal pathway for \ce{CO} is by reaction with \ce{OH}, yielding \ce{CO2} and \ce{H}. The abundance of \ce{HO_x} (\ce{OH} + \ce{HO2}) therefore controls both the recycling of \ce{CO} to \ce{CO2} and the lifetime of \ce{CH4}, which oxidizes through a short sequence that passes through formaldehyde and \ce{CO} before reaching \ce{CO2}. When \ce{HO_x} production is strong, \ce{CO} is efficiently recycled and \ce{CH4} lifetimes are short. When \ce{HO_x} production is weak, \ce{CO} lifetimes increase and \ce{CH4} persists.

We incorporate updated cross sections for \ce{H2O} and \ce{CO2}. Extending \ce{H2O} absorption into the near-UV increases \ce{HO_x} production, speeds conversion of \ce{CO} to \ce{CO2}, and shortens \ce{CH4} lifetimes, which avoids artificial \ce{CO} buildup that can occur if short-wavelength \ce{H2O} absorption is underestimated \citep{ranjanPhotochemistryAnoxicAbiotic2020, broussardImpactExtendedH2O2024}. Updated \ce{CO2} cross sections have a small positive effect on the \ce{CO2} photolysis rates and the partial shielding of \ce{H2O} photolysis \citep{broussardImpactExtendedCO22025}. These updated cross-sections, along with the molecular cross-sections of \ce{CH4}, \ce{O2}, and \ce{O3} are also shown in Figure \ref{fig:stellar_spectra}.

To represent contrasting lower boundary conditions, we prescribe different dry deposition velocities for gases that life is expected to consume efficiently. Following \citet{kharechaCoupledAtmosphereEcosystem2005}, we adopt two deposition velocity regimes. In the biotic deposition velocity cases, strong surface sinks for \ce{CO}, \ce{H2}, and \ce{O2} at $1.2 \times 10^{-4}$, $2.4 \times 10^{-4}$, and $1.4 \times 10^{-4}$ cm s$^{-1}$ are set respectively. Cases with abiotic deposition velocities use weak or negligible sinks at $1.0 \times 10^{-8}$, $0$, and $0$ cm s$^{-1}$, respectively. Cases with abiotic deposition velocities are paired with $F_{\ce{CH4}}$ equal to $10^{8}$, $10^{9}$, and $10^{10}$ molecules cm$^{-2}$ s$^{-1}$, spanning methane fluxes consistent with volcanism up to an ambiguous end member consistent both with a calculated maximum abiotic flux from global serpentinization and with a modest-sized methanogenic biosphere \citep{krissansen-tottonDetectabilityBiosignaturesAnoxic2018}. Cases with biotic deposition velocities are paired with $F_{\ce{CH4}}$ equal to $10^{9}$, $10^{10}$, and $10^{11}$ molecules cm$^{-2}$ s$^{-1}$, representing a flux range constrained by limited \ce{CH4}-producing biospheres up to a modern-Earth-like \ce{CH4} flux \citep{catlingPrebioticEarlyPostbiotic2017, krissansen-tottonDisequilibriumBiosignaturesEarth2018, woganAbundantAtmosphericMethane2020, thompsonCaseContextAtmospheric2022}. The overlapping fluxes are intentionally included in both the abiotic and biotic deposition regimes because all but the highest fluxes are not uniquely diagnostic of either biology or geology. This overlap allows us to isolate how biologically enhanced surface uptake of \ce{CO}, \ce{H2}, and \ce{O2} alters atmospheric chemistry for a fixed \ce{CH4} input.

Lightning is a major natural source of atmospheric \ce{NO_x} on modern Earth \citep{schumannGlobalLightninginducedNitrogen2007}. Our photochemical model includes a lightning source of \ce{NO}, determined by the bulk atmospheric composition and normalized to the lightning production in the atmosphere of modern Earth. \ce{NO} can catalyze the net \ce{CO + O -> CO2} recombination reaction, thereby suppressing photochemical \ce{O2} false positives \citep{harmanAbioticO2Levels2018}. More recent TRAPPIST-1-focused work suggests that substantially higher flash rates may be required to erase abiotic \ce{O3} features, though it explored only one TRAPPIST-1 spectral input, the Peacock+2019 (P19) Median SED \citep{barthEffectLightningAtmospheric2024}.
 
Although three-dimensional models can resolve the latitudinal and longitudinal structure expected for tidally locked worlds such as the TRAPPIST-1 planets, we adopt the one-dimensional photochemical model \texttt{Atmos} because no published three-dimensional photochemical model currently self-consistently simulates terrestrial atmospheres with non-fixed \ce{O2} concentrations in Archean-like conditions. A one-dimensional framework also allows us to explore a broader range of stellar spectra and lower-boundary conditions than would be computationally practical in three dimensions. This approach is therefore well suited for isolating how stellar UV forcing and lower-boundary sinks modulate the coupled \ce{CO2}-\ce{CH4} photo-oxidation system.

\subsection{Synthetic Transmission Spectra Generation with \texttt{SMART}} \label{ssec:transmission_methods}

\begin{figure*}[t!]
\includegraphics[width=\linewidth]{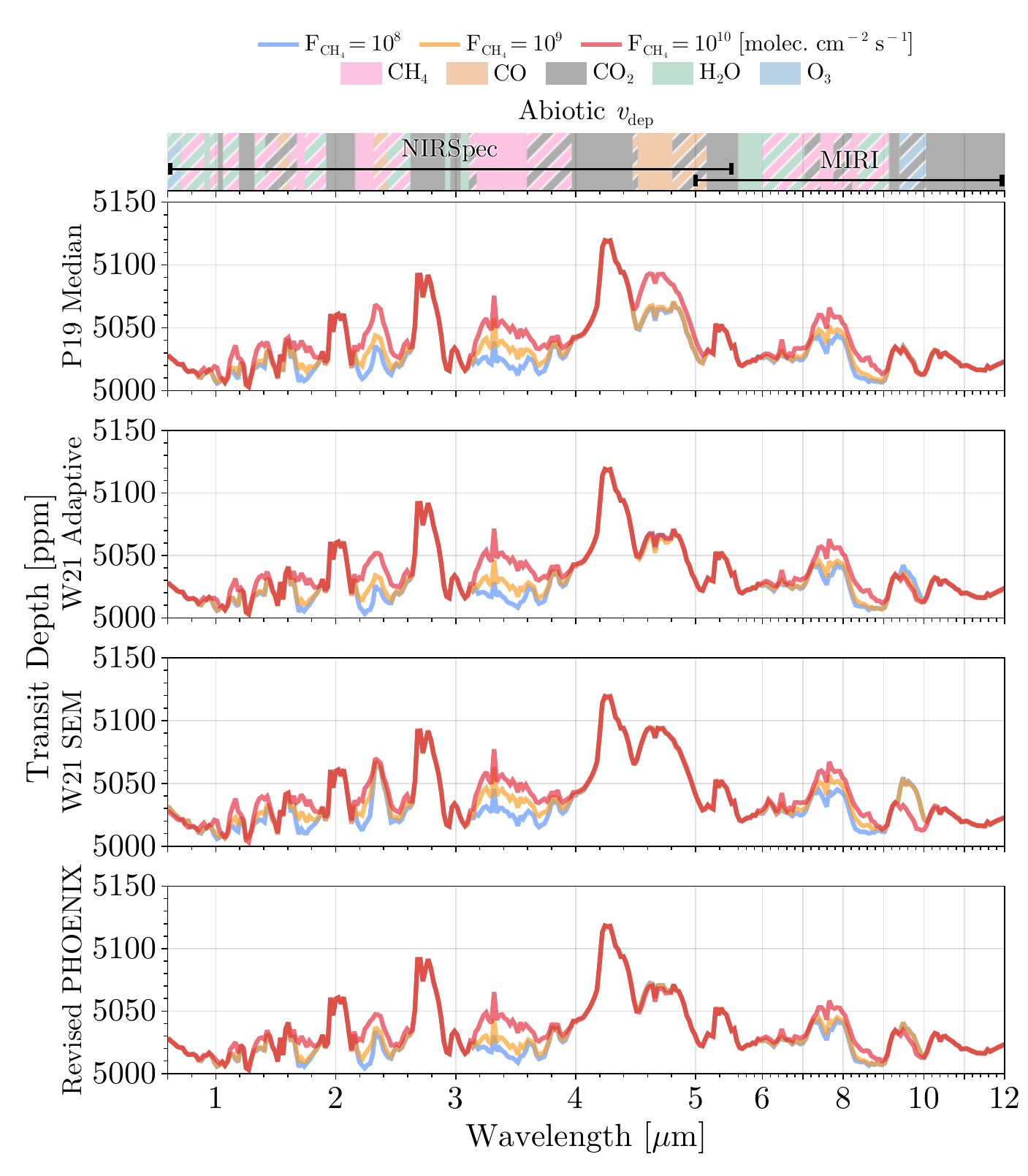}
\caption{Synthetic transmission spectra of Archean-analog TRAPPIST-1 e atmospheres with abiotic deposition velocities generated with \texttt{SMART} for the four adopted TRAPPIST-1 SEDs. Panels spanning the JWST/NIRSpec (\qtyrange{0.6}{5.3}{\micro\meter}) and MIRI (\qtyrange{5.0}{12.0}{\micro\meter}) wavelength ranges are shown for each generated spectrum. Colored regions within the shaded bar indicate broad spectral features of \ce{CH4} (pink), \ce{CO} (orange), \ce{CO2} (gray), \ce{H2O} (green), and \ce{O3} (light blue). Hatched regions indicate wavelength intervals where important spectral features from multiple molecules overlap. Modeled atmospheres with $F_{\ce{CH4}}=10^{8}$, $10^{9}$, and $10^{10}$ molecules cm$^{-2}$ s$^{-1}$ are represented as light blue, yellow, and red lines, respectively. The high-$F_{\ce{CH4}}$ Peacock+2019 (P19) Median case shows a strong \ce{CO} feature at \qty{4.6}{\micro\meter}. Additionally, the transmission spectrum for the two lower \ce{CH4} fluxes in the W21 SEM scenarios display simultaneous \ce{CH4} and \ce{O3} spectral features, though they are accompanied by strong \ce{CO} features at 2.3 and \qty{4.6}{\micro\meter}.}
\label{fig:abiotic_transmission}
\end{figure*}

\begin{figure*}[t!]
\includegraphics[width=\linewidth]{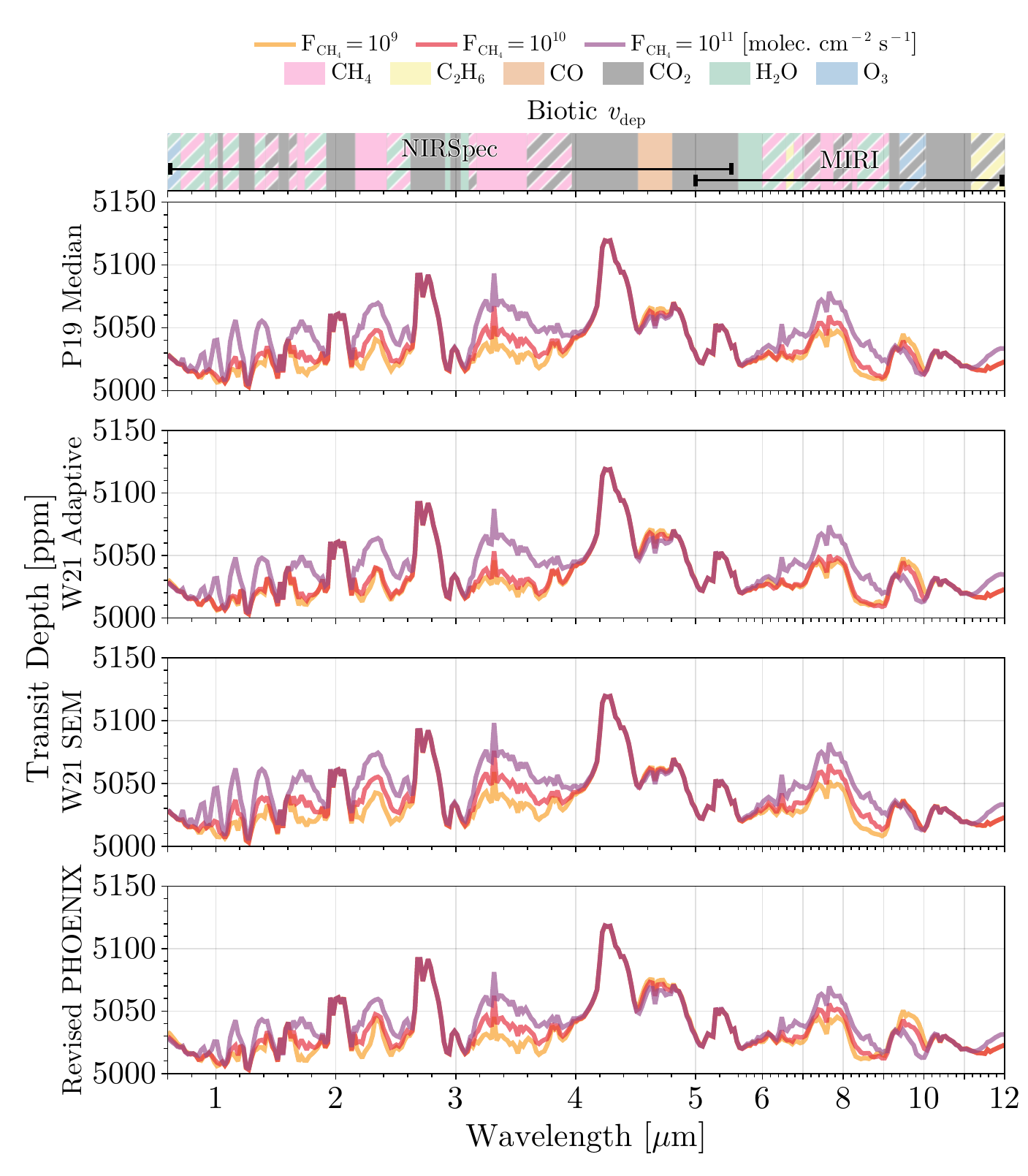}
\caption{Synthetic transmission spectra of Archean-analog TRAPPIST-1 e atmospheres with biotic deposition velocities generated with \texttt{SMART} for the four adopted TRAPPIST-1 SEDs. Panels spanning the JWST/NIRSpec (\qtyrange{0.6}{5.3}{\micro\meter}) and MIRI (\qtyrange{5.0}{12.0}{\micro\meter}) wavelength ranges are shown for each generated spectrum. Colored regions within the shaded bar indicate broad spectral features of \ce{CH4} (pink), \ce{C2H6} (yellow), \ce{CO} (orange), \ce{CO2} (gray), \ce{H2O} (green), and \ce{O3} (light blue). Hatched regions indicate wavelength intervals where important spectral features from multiple molecules overlap. Modeled atmospheres with $F_{\ce{CH4}}=10^{9}$, $10^{10}$, and $10^{11}$ molecules cm$^{-2}$ s$^{-1}$ are represented as yellow, red, and purple lines, respectively. For the Mega-MUSCLES Adaptive case, the nominal $F_{\ce{CH4}}=10^{10}$ molecules cm$^{-2}$ s$^{-1}$ run did not converge in \texttt{Atmos}; we therefore show the nearest converged solution, $F_{\ce{CH4}}=1.5\times10^{10}$ molecules cm$^{-2}$ s$^{-1}$. In the biotic cases, \ce{CO} features are substantially reduced because enhanced surface uptake suppresses atmospheric \ce{CO}, while in the MIRI wavelength range several low-$F_{\ce{CH4}}$ cases retain appreciable \ce{O3} absorption near \qty{9.7}{\micro\meter}, though this feature weakens as $F_{\ce{CH4}}$ increases.}
\label{fig:biotic_transmission}
\end{figure*}

For each individual \texttt{Atmos} run, we generate atmospheric profiles and then use them to compute synthetic transmission spectra that include atmospheric refraction but neglect clouds.
We use the \texttt{Spectral Mapping and Atmospheric Radiative Transfer} (\texttt{SMART}) one-dimensional line-by-line, multi-stream, multiple-scattering radiative transfer model to generate the synthetic spectra \citep{stamnesNumericallyStableAlgorithm1988,meadowsGroundbasedNearinfraredObservations1996, crispAbsorptionSunlightWater1997}. \texttt{SMART} uses line lists sourced from the high-resolution transmission molecular absorption database (HITRAN-2020) \citep{gordonHITRAN2020MolecularSpectroscopic2022} to calculate the opacities of major atmospheric species using its companion program \texttt{LBLABC}.
The resulting synthetic transmission spectra are shown in Figures~\ref{fig:abiotic_transmission} and \ref{fig:biotic_transmission}. Each figure includes both the JWST/NIRSpec (\qtyrange{0.6}{5.3}{\micro\meter}) and MIRI (\qtyrange{5.0}{12.0}{\micro\meter}) wavelength ranges, and the plotted spectra are down-binned to match the resolutions of the corresponding instrument modes. Although we primarily focus on the implications for interpreting space-based infrared transmission spectra with JWST or successor telescope concepts such as the Origins Space Telescope \citep{battersbyOriginsSpaceTelescope2018, meixnerOriginsSpaceTelescope2019, leisawitzOriginsSpaceTelescope2021}, these wavelength ranges are also expected to be observable with ground-based ELT-class telescopes \citep{hardegree-ullmanBioverseComprehensiveAssessment2023, currieTheresMoreLife2023}.

\section{Results}\label{sec:results}

\subsection{Photochemical Results} \label{ssec:photochemical_results}

For each of the four TRAPPIST-1 SEDs, we run six \texttt{Atmos} photochemical simulations: three with abiotic deposition velocities and three with biotic deposition velocities, each paired with the corresponding prescribed \ce{CH4} surface fluxes. These results are plotted in Figure~\ref{fig:spaghetti}. One case requires a small exception: for the biotic Mega-MUSCLES Adaptive scenario, the nominal $F_{\ce{CH4}}=10^{10}$ molecules cm$^{-2}$ s$^{-1}$ run did not converge in \texttt{Atmos}, so we instead show the nearest converged solution at $F_{\ce{CH4}}=1.5\times10^{10}$ molecules cm$^{-2}$ s$^{-1}$.

Increasing \ce{CH4} surface fluxes increases the \ce{CH4} mixing ratios as expected for all SEDs considered. For any fixed \ce{CH4} surface flux, however, different SEDs produce order-of-magnitude variations in the resulting \ce{CH4} abundance. Across our parameter space explored, the \ce{CH4} mixing ratios span nearly three orders of magnitude, set jointly by the input SED, the imposed \ce{CH4} surface flux, and the choice of abiotic versus biotic deposition velocities.

For most of the modeled cases with low \ce{CH4} fluxes, \ce{CO} mixing ratios account for only a small percentage of the atmosphere. However, as \ce{CH4} fluxes increase, there is more \ce{CH4} available to be photo-oxidized to \ce{CO}.
Although this \ce{CO} can be further oxidized to \ce{CO2}, this reaction is rate-limited, as there is not enough \ce{OH} available, and thus \ce{CO} is produced faster than it can be removed \citep{kastingPhotochemistryMethaneEarths1983, kharechaCoupledAtmosphereEcosystem2005, zahnlePhotochemicalInstabilityAncient2008,ranjanPhotochemicalRunawayExoplanet2022}. As a result, \ce{CO} can accumulate in the atmospheres modeled under the abiotic regime (i.e., in the absence of robust surface sinks), particularly in the \enquote{Peacock+2019 Median} and \enquote{Mega-MUSCLES SEM} SEDs. However, the mixing ratios of \ce{CO} are kept three to four orders of magnitude lower in the cases with biotic deposition velocities compared to those with abiotic deposition velocities, since more \ce{CO} is removed from the atmosphere as the surface deposition velocity increases.

\ce{O2} and \ce{O3} abundances vary strongly across the modeled cases, with their significance being strongly spectrum- and condition-dependent. The clearest abiotic oxidant buildup occurs in the low-\ce{CH4}-flux Mega-MUSCLES~SEM case, where the surface \ce{O2} mixing ratio reaches 5.3\%. The biotic runs, especially for low \ce{CH4} fluxes, can also build up \ce{O2}; for example, the \enquote{Revised PHOENIX} biotic case has a surface \ce{O2} mixing ratio of about 390 ppm. In these oxidized regimes, \ce{O2} accumulates when O atoms generated by \ce{CO2} photolysis under FUV irradiation recombine through three-body reactions such as \ce{O + O + M -> O2 + M}, where \ce{M} denotes a background atmospheric molecule that carries away excess energy, while reductant inputs remain low \citep{harmanAbioticO2Levels2015}. In the biotic cases, enhanced surface uptake of \ce{CO} further promotes oxidant buildup by suppressing the dominant \ce{CO + OH} recombination pathway, thereby allowing more photochemically produced oxygen to persist as \ce{O2} and, downstream, \ce{O3}, consistent with the mechanism identified by \citet{eager-nashSimulatingBiosignaturesPreoxygen2024}.

We also note that the more oxidized runs show enhanced \ce{HNO3}, indicating more active \ce{NO_x} cycling in the atmosphere. The ozone-loss step of interest is \ce{NO + O3 -> NO2 + O2}, while enhanced \ce{HNO3} reflects downstream \ce{NO_x} processing through reactions such as \ce{NO2 + OH + M -> HNO3 + M}. To assess the importance of this chemistry more directly, Appendix~\ref{app:o3_loss} quantifies the fraction of total \ce{O3} loss attributable to \ce{NO + O3} for our prescribed \ce{NO} source. Across the SED suite, this pathway is negligible in most of the cases with abiotic deposition velocities, but becomes modestly important in several low- and intermediate-$F_{\ce{CH4}}$ biotic runs, consistent with reduced atmospheric \ce{CO} leaving more \ce{OH} available for \ce{NO_x} processing rather than being consumed by \ce{CO + OH -> CO2 + H}. The Mega-MUSCLES~SEM cases are the main exception: the run with abiotic deposition velocities already shows a non-negligible \ce{NO + O3} contribution, whereas the corresponding biotic run exhibits a substantially stronger \ce{NO_x}-mediated ozone sink. Thus, although the low-$F_{\ce{CH4}}$ Mega-MUSCLES~SEM case with abiotic deposition velocities can sustain large \ce{O2} and \ce{O3} abundances, the corresponding biotic case is more effectively limited by the combination of an explicit lower-boundary \ce{O2} sink and stronger \ce{NO_x}-mediated \ce{O3} loss.

\subsection{Synthetic Transmission Spectra} \label{ssec:transmission_results}

Each converged photochemical run was also used as input to \texttt{SMART} to generate synthetic cloud-free transmission spectra. Figure~\ref{fig:abiotic_transmission} shows the cases with abiotic deposition velocities, while Figure~\ref{fig:biotic_transmission} shows the cases with biotic deposition velocities, each spanning both the JWST/NIRSpec and MIRI wavelength ranges. We adopt cloud-free spectra as an idealized baseline to isolate the spectral differences between the photochemical solutions. Low-altitude \ce{H2O} clouds could reduce absolute feature amplitudes, especially for lower-atmosphere absorbers, but would not necessarily obscure the \ce{CH4}, \ce{CO}, \ce{CO2}, and \ce{O3} bands emphasized here, as their effective transit depths lie above the cloud deck. These spectra are therefore best interpreted as cloud-free reference cases rather than cloud-inclusive detectability predictions \citep{fauchezImpactCloudsHazes2019}.

\begin{figure*}[t!]
    \centering
    \includegraphics[width=\linewidth]{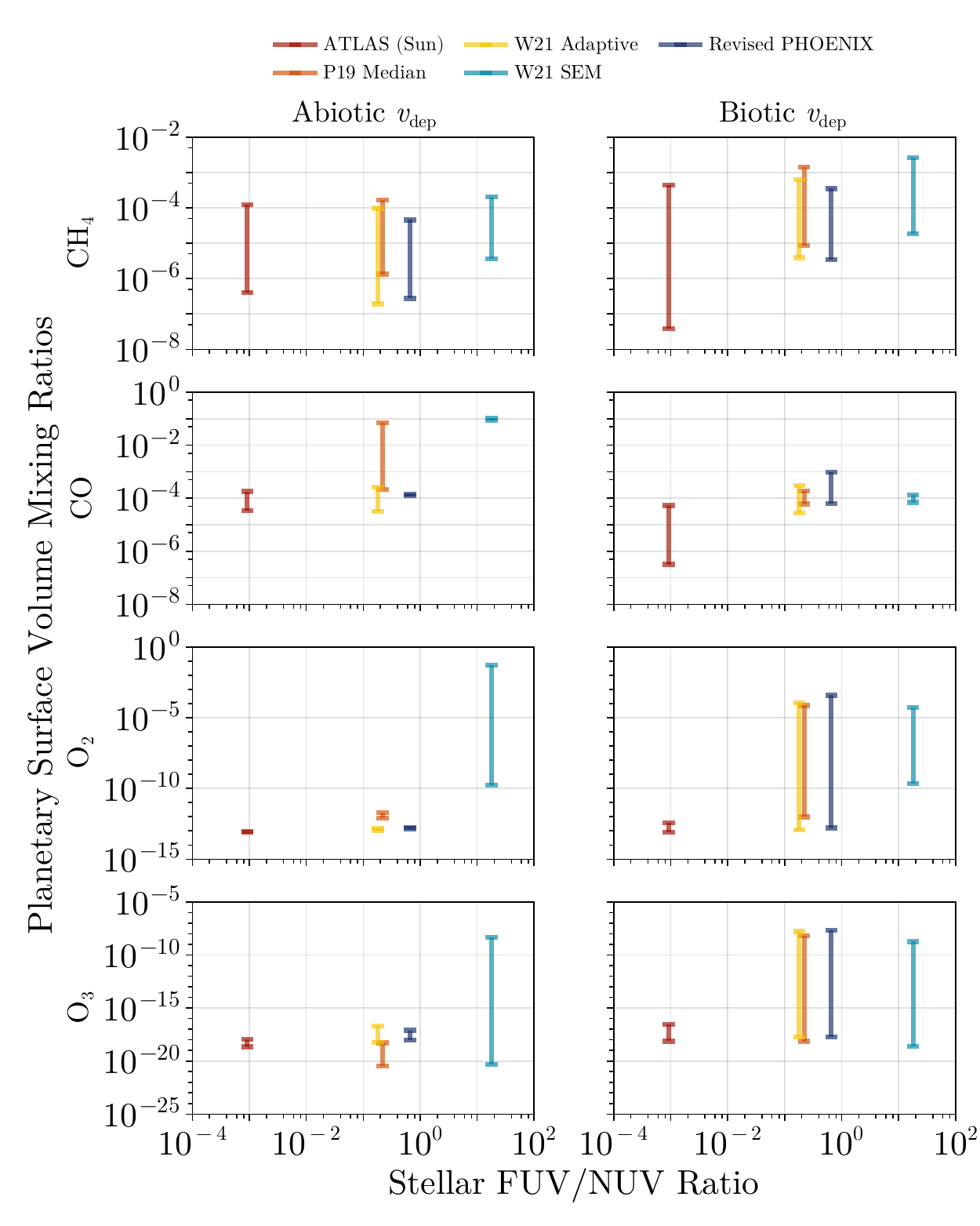}
    \caption{The range of planetary surface volume mixing ratios of \ce{CH4}, \ce{CO}, \ce{O2} and \ce{O3} are plotted as a function of the FUV/NUV ratio, where the far-ultraviolet is defined from \num{120.0} to \qty{175.0}{\nano\meter} and the near-ultraviolet is defined from \num{175.0} to \qty{312.5}{\nano\meter}. Whiskers span the three imposed methane surface fluxes for each of the four TRAPPIST-1 spectral energy distributions as well as a scaled version of the modern Sun's spectrum \citep{thuillierSolarIrradianceReference2004}. Cases with abiotic deposition velocities, using \ce{CH4} fluxes of $10^8$, $10^9$, and $10^{10}$ molecules cm$^{-2}$ s$^{-1}$ are plotted on the left and cases with biotic deposition velocities, using \ce{CH4} fluxes of $10^9$, $10^{10}$, and $10^{11}$ molecules cm$^{-2}$ s$^{-1}$, are plotted on the right.} 
    \label{fig:fuvnuv}
\end{figure*}

Across both figures, increasing the imposed \ce{CH4} surface flux produces systematically stronger \ce{CH4} absorption features. At fixed $F_{\ce{CH4}}$, spectra with larger FUV/NUV ratios, notably the Mega-MUSCLES~SEM SED, yield the strongest methane features (see Figure~\ref{fig:fuvnuv}). In these cases, the weak NUV flux suppresses \ce{HO_x} production, extending the atmospheric lifetime of \ce{CH4}. Conversely, the other SEDs, with stronger NUV fluxes, generate more \ce{HO_x}, allowing more rapid oxidation of \ce{CH4}, decreasing its column abundance and muting the associated spectral features. This behavior is most apparent in the broad \qty{3.3}{\micro\meter} and \qty{7.7}{\micro\meter} bands, although several narrower \ce{CH4} features across both wavelength ranges also vary appreciably.

In the spectra corresponding to cases with abiotic deposition velocities (Figure~\ref{fig:abiotic_transmission}), \ce{CO} features become increasingly apparent at high \ce{CH4} surface fluxes. This effect is most noticeable in the \qtyrange{4.6}{4.8}{\micro\meter} band and reflects enhanced atmospheric \ce{CO} accumulation under weak surface uptake (see Figure~\ref{fig:spaghetti}). A distinct \ce{O3} feature near \qty{9.7}{\micro\meter} is also apparent and is most evident in the low-$F_{\ce{CH4}}$ Mega-MUSCLES~SEM case, while the other cases with abiotic deposition velocities show only weak or absent \ce{O3} features. As a result, for one TRAPPIST-1 UV reconstruction, an atmosphere can sustain simultaneous spectral signatures of both \ce{CH4} and \ce{O3} without life.

In the spectra generated from the cases with biotic deposition velocities (Figure~\ref{fig:biotic_transmission}), \ce{CO} absorption is substantially reduced relative to the cases with abiotic deposition velocities because enhanced surface uptake suppresses atmospheric \ce{CO}. At the same time, several low-$F_{\ce{CH4}}$ biotic cases retain appreciable \ce{O3} absorption near \qty{9.7}{\micro\meter}, although this feature weakens substantially as $F_{\ce{CH4}}$ increases and the atmosphere becomes more reducing. These spectra therefore highlight the contrasting observational outcomes of biologically enhanced \ce{CO} consumption: muted \ce{CO} features together with retained oxidant signatures.

Although organic haze is typically considered to hide other spectral features, it can also act as a biosignature when present in high-\ce{CO2} atmospheres, as a high surface flux of \ce{CH4} is necessary to produce these haze layers. However, since the mixing ratio of \ce{CH4} never rises above a tenth of the mixing ratio of \ce{CO2}, the formation of haze is limited, and thus significant haze features are not present within these spectra \citep{trainerOrganicHazeTitan2006, arneyPaleOrangeDots2017}. 

\section{Discussion} \label{sec:discussion}

Our results demonstrate that uncertainties in TRAPPIST-1's UV SEDs propagate directly into the chemical context in which biosignatures are interpreted. For Archean-analog atmospheres around ultracool M dwarfs, the main ambiguity is not the \ce{CH4}-\ce{CO2} disequilibrium framework by itself, which is typically considered alongside low \ce{CO}, but whether the combined \ce{CH4}-\ce{CO2}-\ce{O3} signature can arise abiotically. For some UV characterizations of TRAPPIST-1, including the Mega-MUSCLES~SEM, abiotic and biotic deposition-velocity scenarios can produce overlapping \ce{O2} and \ce{O3} abundance regimes. In such cases, the coexistence of \ce{CH4}, \ce{CO2}, and \ce{O3} is not by itself uniquely diagnostic of life. Instead, \ce{CO} provides critical context: abundant \ce{CO} points to abiotic \ce{O3} accumulation driven by \ce{CO2} photolysis and inefficient surface removal, whereas muted \ce{CO} at comparable \ce{CH4} and \ce{O3} levels is more consistent with efficient biological \ce{CO} consumption.

Two major pathways capture the accumulation of \ce{O2} and \ce{O3} within these Archean-analog atmospheres. First, in the cases with abiotic deposition velocities, the high FUV/NUV ratio associated with the Mega-MUSCLES~SEM SED produces elevated \ce{O2} at low \ce{CH4} surface fluxes because the enhanced FUV flux drives \ce{CO2} photolysis and supplies free oxygen atoms while the NUV flux is too weak to sustain strong photolysis of \ce{H2O}, which would produce radicals that could catalyze the destruction of \ce{O3} and \ce{CH4}. Second, in biotic cases, elevated \ce{CO} dry deposition velocities suppress atmospheric \ce{CO} and thus weaken the dominant \ce{CO + OH} recombination pathway. Under low-\ce{CH4}-flux conditions, this combination of FUV-driven oxidant production and biotically enhanced \ce{CO} sinks allows both \ce{O2} and \ce{O3} to accumulate. In these scenarios, \ce{O3} could still be biologically mediated in the sense that biological metabolisms help maintain the \ce{CO} sinks that allow oxidants to accumulate, but the life responsible for this oxidation does not need to employ oxygenic photosynthesis.

\subsection{Degeneracies Controlled by the Stellar FUV/NUV Ratio}

Figure~\ref{fig:fuvnuv} summarizes these behaviors using the stellar FUV/NUV ratio as a convenient proxy for differences among the input spectra. Although the FUV/NUV ratio is not a complete predictor of the resulting atmospheric state, it represents the dominant contrast among the adopted SEDs: \enquote{Mega-MUSCLES SEM} combines strong short-wavelength oxidant production with relatively weak longer-wavelength radical and ozone-loss pathways, producing the most oxidized abiotic solutions in our model suite. The spread in whisker ranges therefore highlights that changes in the assumed TRAPPIST-1 UV reconstruction can move Archean-analog atmospheres between relatively weakly and strongly oxidized states. In the cases with biotic deposition velocities, \ce{O2} and \ce{O3} mixing ratios also span many orders of magnitude, but their behavior reflects the combined effects of stellar UV forcing, imposed \ce{CH4} flux, enhanced \ce{CO} uptake, the lower-boundary \ce{O2} sink, and \ce{NO_x}-mediated ozone loss. This behavior is most visually distinct in the cases with abiotic deposition velocities, where \enquote{Mega-MUSCLES SEM} occupies a high-FUV/NUV, oxidant-rich regime. Because the abiotic and biotic panels use \ce{CH4} ranges offset by an order of magnitude, the ranges should be interpreted as regime-specific envelopes rather than as a one-to-one comparison. A similar figure can be found in Appendix~\ref{app:fuvnuv_columns} that presents the same comparison using atmospheric column densities instead of planetary surface volume mixing ratios.

These trends are consistent with earlier work showing that high FUV/NUV ratios can drive \ce{CO2}-dominated atmospheres toward more oxidized states by enhancing \ce{CO2} photolysis relative to \ce{H2O} and reduced gases \citep{domagal-goldmanAbioticOzoneOxygen2014, grenfellSensitivityBiosignaturesEarthlike2014, harmanAbioticO2Levels2015, kozakisOzoneReliableProxy2022, tianHighStellarFUV2014}. Coupled atmosphere-biogeochemistry models have likewise found that M dwarf planets can be easier to oxygenate than their counterparts orbiting Sun-like stars for a given net primary productivity because their UV environments favor net oxidation \citep{gebauerEvolutionEarthlikePlanetary2018}. In this context, TRAPPIST-1 is an end-member: depending on which UV reconstruction is adopted, the system may occupy either a regime that favors strong photochemical oxidation and abiotic \ce{O2}/\ce{O3} buildup or a more weakly oxidizing regime. The present uncertainty in TRAPPIST-1’s true FUV-NUV spectrum therefore underpins the plausibility of both the abiotic false positives and the non-photosynthetic, oxygen-rich biospheres recovered in our Archean-analog models.

\subsection{Atmospheric Oxygenation}

Our biotic runs, which include enhanced \ce{CO}, \ce{H2}, and \ce{O2} deposition velocities to represent \ce{CO}-, \ce{H2}-, and \ce{O2}-consuming metabolisms, can produce percent-level \ce{O2} surface mixing ratios together with substantial \ce{O3} columns. This result closely parallels the coupled one-dimensional atmosphere-ocean-ecosystem simulations of \citet{eager-nashSimulatingBiosignaturesPreoxygen2024}, who showed that \ce{CO}-consuming biospheres on TRAPPIST-1 e can oxygenate an atmosphere by suppressing atmospheric \ce{CO} and thereby weakening major oxygen-loss pathways. In our framework, we recover the same qualitative oxygenation mechanism but additionally show that the magnitude of the associated \ce{O3} buildup, and thus the strength of the resulting \ce{O3} spectral signature, depends strongly on the adopted stellar spectrum. As a result, similar biotic \ce{O2} states can map onto different \ce{O3} outcomes across our SED suite, and in some cases, those biotic oxidant states overlap with purely abiotic, stellar-UV-driven solutions, particularly for Mega-MUSCLES~SEM.

A key implication is that atmospheric oxygenation is fundamentally a statement about net redox balance, rather than oxygenic photosynthesis per se. On Earth, the rise of \ce{O2} reflects the long-term imbalance between organic matter production and oxidative recycling, with burial of reduced carbon providing the net source of atmospheric \ce{O2} \citep{lyonsRiseOxygenEarth2014, catlingRiseOxygenOzone2017}. Stoichiometrically, both oxygenic photosynthesis followed by organic carbon burial and anoxygenic metabolisms that consume \ce{CO} to produce organic matter while exporting reduced carbon can, in principle, drive the atmosphere toward higher \ce{O2}. Our model does not include an explicit time-dependent carbon cycle, but \citet{eager-nashSimulatingBiosignaturesPreoxygen2024} already took an important step in this direction with a coupled atmosphere-ocean-ecosystem framework that includes redox balance and organic carbon burial. Extending such coupled models, and connecting them more explicitly to broader M dwarf atmosphere-biogeochemical frameworks such as in \citet{gebauerEvolutionEarthlikePlanetary2018}, remains an important next step.

More broadly, the mechanism identified here is not unique to TRAPPIST-1. Any system in which anoxygenic metabolisms efficiently reduce carbon and export or bury that carbon could, in principle, sustain an \ce{O2}-\ce{O3}-rich atmosphere without oxygenic photosynthesis. At the same time, our results show that around ultracool M dwarfs the interpretability of such oxidized states depends strongly on the stellar UV spectrum, because the same biological oxygenation pathway can produce markedly different \ce{O3} abundances and spectral signatures under different SEDs. Distinguishing such biospheres from abiotic false positives will therefore remain difficult: both can yield overlapping \ce{O2}/\ce{O3} abundances and similar \ce{CH4}-\ce{CO}-\ce{O2}-\ce{O3} combinations. In this regime, oxidant abundance alone is not a unique tracer of oxygenic photosynthesis, and interpretation will require the combined context of stellar UV characterization and planetary redox state.

\subsection{Observational Implications}

Our synthetic transmission spectra show that these photochemical differences map directly onto observables. Across all SEDs, the input stellar spectrum alters the modeled photochemistry and, in turn, the resulting transmission features. Increased $F_{\ce{CH4}}$ generally produces larger atmospheric \ce{CH4} abundances and therefore stronger \ce{CH4} absorption bands in both the NIRSpec (\qtyrange{0.6}{5.3}{\micro\meter}) and MIRI (\qtyrange{5}{12}{\micro\meter}) wavelength ranges. At fixed $F_{\ce{CH4}}$, band depths depend strongly on the adopted stellar SED, because lower NUV fluxes suppress \ce{HO_x} production from \ce{H2O} photolysis, thereby extending the atmospheric lifetime of \ce{CH4}. The same SED-dependent photochemistry also affects the observability of \ce{CO} in the NIRSpec range and \ce{O3} near \qty{9.7}{\micro\meter} in the MIRI range. Strong \ce{CO} bands co-occurring with \ce{CH4} features indicate slow \ce{CO} oxidation and are characteristic of some of our lifeless cases, particularly those that accumulate spectrally discernible \ce{O3}. In contrast, enhanced microbially-mediated surface uptake of \ce{CO} suppresses the \ce{CO} spectral features while permitting \ce{O2} and \ce{O3} to build up, yielding spectra in which strong \ce{CH4} and \ce{O3} features coexist despite low \ce{CO}. Across our model scenarios, \ce{CO} therefore retains its value as an anti-biosignature: lifeless cases show strong \ce{CO} bands, which become especially prominent at the highest \ce{CH4} fluxes, whereas biotic cases show muted \ce{CO} bands at comparable \ce{CH4} levels (though we note that assuming maximum aqueous biotic deposition rates may be optimistic, and ambiguous high-\ce{CO} cases with life could arise \citep[e.g.,][]{schwietermanRethinkingCOAntibiosignatures2019}).

In practice, separating these scenarios will be challenging with JWST alone. Detecting \ce{O3} in transmission around ultracool M dwarfs is difficult even under optimistic noise assumptions, and both stellar contamination and instrument systematics can further mute small features. Thus, as previous studies have emphasized, both abiotic \ce{O2}/\ce{O3} production pathways and the modest signal amplitudes expected for realistic TRAPPIST-1 e observations complicate robust \ce{O3} detection and interpretation with JWST \citep{meadowsReflectionsO2Biosignature2017, meadowsExoplanetBiosignaturesUnderstanding2018, meadowsFeasibilityDetectingBiosignatures2023}. Our spectra should therefore be viewed as optimistic, noise-free upper bounds. We do not present a formal noise analysis here and therefore do not quantify retrieval significance on an instrument-by-instrument basis. Many of the spectral differences identified in this work would likely require numerous transits to detect with confidence, especially for weaker \ce{CO} and \ce{O3} features. In real data, many of the atmospheric degeneracies we identify are likely to be only partially resolved with current capabilities. We likely require future platforms such as those like the proposed Origins Space Telescope \citep{battersbyOriginsSpaceTelescope2018, meixnerOriginsSpaceTelescope2019, leisawitzOriginsSpaceTelescope2021} or yet more advanced instrumentation to better resolve these spectral differences.

\subsection{Model Limitations}

Our results come from a one-dimensional, steady-state photochemical model with prescribed lower-boundary fluxes and a fixed surface pressure. These assumptions are appropriate for isolating the effects of stellar spectral uncertainty, but they also introduce important limitations. In particular, this study is designed to test the sensitivity of Archean-analog TRAPPIST-1 e photochemistry to different plausible stellar UV spectra, rather than to provide a fully coupled treatment of climate, biogeochemistry, interior exchange, and spatial transport.

These limitations are relevant for tidally locked planets such as TRAPPIST-1 e, where three-dimensional circulation, spatially heterogeneous photolysis, and day-night transport could alter the relationship between local chemical production and disk-integrated observables. In addition, our lower-boundary conditions prescribe biological and abiotic surface sinks rather than calculating them from an explicit ocean-atmosphere-ecosystem model. Coupled atmosphere-ocean-ecosystem and atmosphere-biogeochemical models have already shown that redox balance, biological boundary conditions, and M dwarf spectral forcing can strongly modulate atmospheric \ce{O2} production, loss, and net oxygenation \citep{gebauerEvolutionEarthlikePlanetary2018, eager-nashSimulatingBiosignaturesPreoxygen2024}. Our results should therefore be interpreted as controlled sensitivity experiments that isolate the consequences of stellar UV uncertainty, rather than as complete predictions of any single TRAPPIST-1 e atmospheric state.

\section{Conclusion} \label{sec:conclusion}

Our results indicate that biosignature interpretation for Archean-analog atmospheres around TRAPPIST-1 e depends strongly on the adopted stellar UV SED. Across otherwise identical planetary boundary conditions, different plausible TRAPPIST-1 UV reconstructions produce large differences in atmospheric \ce{CH4}, \ce{CO}, \ce{O2}, and \ce{O3} abundances. These differences map directly onto synthetic transmission spectra. In particular, some UV prescriptions of TRAPPIST-1 allow \ce{CH4} and \ce{O3} to coexist in atmospheres without oxygenic photosynthesis, including a case where \ce{O3} is produced through abiotic \ce{CO2} photolysis and inefficient radical-driven loss while \ce{CH4} accumulates from fluxes consistent with volcanic outgassing. This creates a potential \ce{CH4}-\ce{O3} false positive for life if the stellar UV environment is not sufficiently well constrained.

In this regime, \ce{CO} becomes one of the most important contextual gases. In our models, abundant \ce{CO} indicates abiotic oxidant accumulation driven by \ce{CO2} photolysis and inefficient surface removal, whereas muted \ce{CO} at comparable \ce{CH4} and \ce{O3} levels is more consistent with efficient biological \ce{CO} consumption. \ce{CO} therefore remains a critical anti-biosignature, when remotely detectable, for distinguishing lifeless high \ce{CO} atmospheres from biologically mediated oxidized states \citep{wangDetectionCarbonMonoxide2016, schwietermanIdentifyingPlanetaryBiosignature2016, watanabeRelativeAbundancesCO22024, woganWhenChemicalDisequilibrium2020}. (However, a more subtle point is that inefficient biological \ce{CO} consumption, potentially coupled with direct and indirect biological \ce{CO} production, could mimic abiotic \ce{CO} accumulation within the atmosphere, resulting in ambiguous scenarios and potential false negatives for life \citep{schwietermanRethinkingCOAntibiosignatures2019, molinaExploringExtremophileGas2026, zhanOrganicCarbonylsAre2022}.) More broadly, our results reinforce that biosignature gases must be interpreted within the broader context of the atmosphere-surface-stellar system rather than as isolated pairwise detections.

A central implication of this work is that stellar UV characterization is a prerequisite for accurately modeling temperate planets orbiting ultracool M dwarf stars. Better empirical constraints on the FUV and NUV spectra of TRAPPIST-1 and similar stars are needed to determine which photochemical regimes are plausible for observed planets. Future coupled photochemical, climate, interior-exchange, and atmosphere-ocean-ecosystem models will also be needed to better connect stellar UV forcing to potentially observable atmospheric spectra. Extending this spectral-sensitivity framework to other nearby habitable-zone M dwarf planets, including Teegarden's Star b and c, LHS 1140 b, and Proxima Centauri b, would further clarify how stellar-spectrum uncertainty shapes atmospheric interpretation beyond the TRAPPIST-1 system. Such work should also assess the impact of stellar variability, including flares, alongside sensitivity to continuing laboratory updates to photolysis cross sections and reaction rates \citep{seguraEffectStrongStellar2010, miranda-roseteBiosignatureFalsePositives2025}. For Archean-analog planets around the coolest M dwarfs, the strongest biosignature interpretations will therefore come not from any single gas or gas pair but from the combined context of the stellar UV environment and detailed atmospheric characterization, including discerning the presence of atmospheric disequilibria and potential anti-biosignature gases.

\begin{acknowledgments}
ELS and EWS thank C. E. Harman for helping to calculate the wavelength ratios used in this manuscript and M. Leung for helping to implement the modeling framework. ELS additionally thanks T. Fetherolf, R. C. Payne, and M. L. Wong for their personal support and mentorship. This work was supported by the NASA Interdisciplinary Consortia for Astrobiology Research (ICAR) Program via the Alternative Earths team, with funding issued under grant No. 80NSSC21K0594 and the Consortium on Habitability and Atmospheres of M-dwarf Planets (CHAMPS) team, with funding issued under grant No. 80NSSC23K1399. ELS, EWS, and NFW acknowledge support from the Virtual Planetary Laboratory, a member of the NASA Nexus for Exoplanet System Science (NExSS), funded via the NASA Astrobiology Program grant No. 80NSSC23K1398. SRP also acknowledges support from NASA under award number 80GSFC24M0006. Computations were performed using the computer clusters and data storage resources of University of California, Riverside's High-Performance Computing Center, which was funded by grants from NSF (MRI-2215705, MRI-1429826) and NIH (1S10OD016290-01A1).

\end{acknowledgments}

\begin{contribution}

ELS led the project, performed the modeling, developed code, generated figures, and wrote the manuscript. EWS and TWL conceptualized the project, provided supervision, and secured funding. SRP contributed a stellar input spectrum used in the simulations. NFW contributed relevant scientific expertise to the project. All authors contributed to the interpretation of results and to the editing of the manuscript.

\end{contribution}

\software{\texttt{Astropy} \citep{robitailleAstropyCommunityPython2013, collaborationAstropyProjectBuilding2018, collaborationAstropyProjectSustaining2022}, \texttt{Atmos} \citep{arneyPaleOrangeDot2016}, \texttt{Julia} \citep{bezansonJuliaFreshApproach2017}, \texttt{Makie} \citep{danischMakieJlFlexible2021}, \texttt{NumPy} \citep{harrisArrayProgrammingNumPy2020}, \texttt{Python} \citep{vanrossumInteractivelyTestingRemote1991}, \texttt{SMART} \citep{meadowsGroundbasedNearinfraredObservations1996, crispAbsorptionSunlightWater1997}, \texttt{SpectRes} \citep{carnallSpectResFastSpectral2017}}

\appendix

\section{Nitric Oxide-Mediated Ozone Loss} \label{app:o3_loss}

\ce{NO_x} chemistry is a potentially important control on oxidant buildup in \ce{CO2}-rich atmospheres because \ce{NO} can participate in catalytic cycles that promote the recombination of \ce{CO} and free oxygen. Previous work has shown that lightning-produced \ce{NO} can suppress the photochemical buildup of \ce{O2} and \ce{O3} in high FUV environments \citep{harmanAbioticO2Levels2018}. \citet{barthEffectLightningAtmospheric2024} suggest that for TRAPPIST-1, substantially enhanced lightning activity may be required to eliminate abiotic ozone features, but they considered only a single TRAPPIST-1 spectral input constructed as the median of the three models from \citet{peacockPredictingExtremeUltraviolet2019}.

\begin{figure*}[b!]
    \centering
    \includegraphics[width=\linewidth]{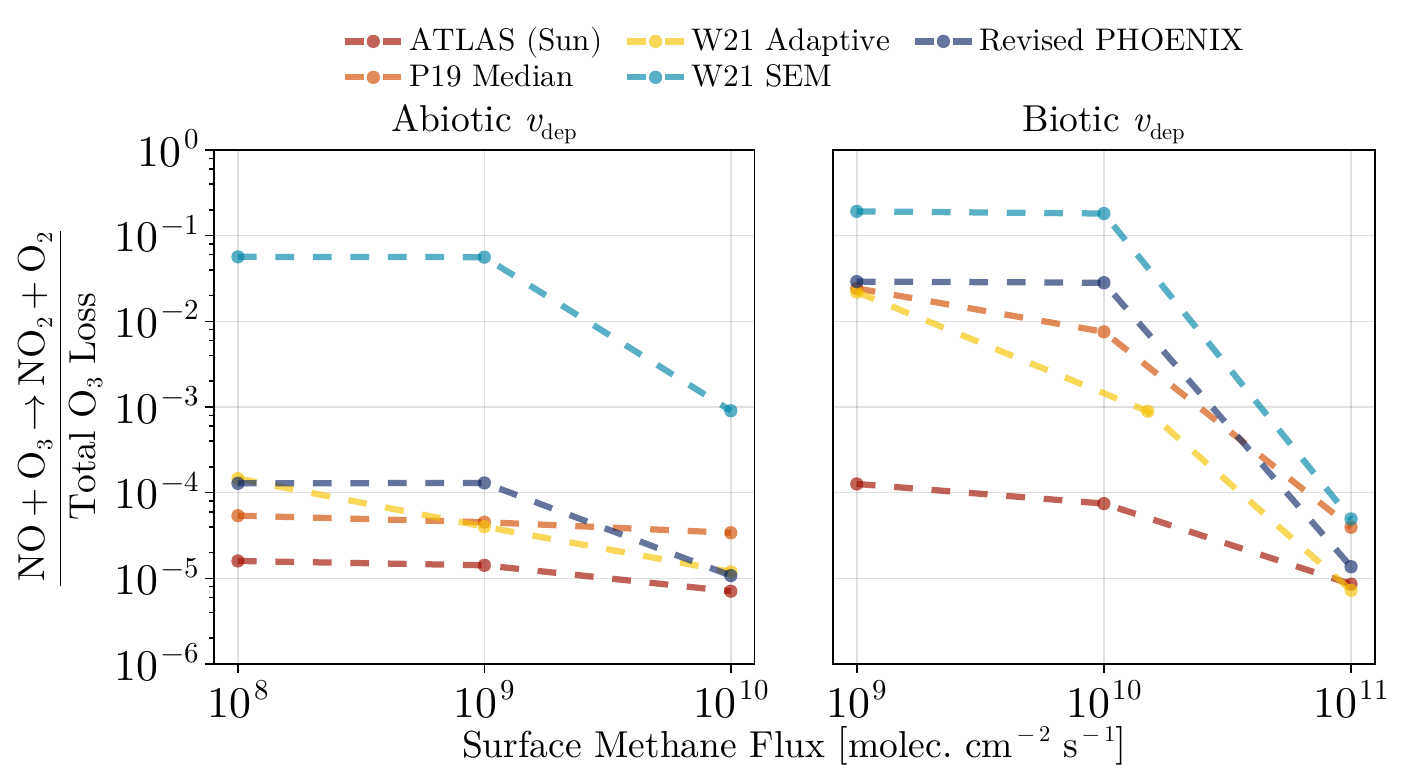}
    \caption{Fraction of total \ce{O3} loss attributable to the \ce{NO + O3} reaction as a function of imposed methane surface flux, shown for abiotic (left) and biotic (right) deposition velocities. The abiotic panel uses $F_{\ce{CH4}}$ equal to $10^{8}$, $10^{9}$, and $10^{10}$ molecules cm$^{-2}$ s$^{-1}$, while the biotic panel uses $F_{\ce{CH4}}$ equal to $10^{9}$, $10^{10}$, and $10^{11}$ molecules cm$^{-2}$ s$^{-1}$. Colored lines correspond to a scaled modern solar spectrum and the four SEDs considered for TRAPPIST-1. The contribution of \ce{NO + O3} to total ozone loss is set not only by the prescribed \ce{NO} source but also by the stellar-SED-dependent photochemical environment. It is generally limited in most of the cases with abiotic deposition velocities but becomes a non-negligible ozone sink in several low- and intermediate-$F_{\ce{CH4}}$ cases with biotic deposition velocities. In both regimes, this pathway is most important for scenarios using the Mega-MUSCLES (W21) SEM input stellar spectrum.}
    \label{fig:no_o3_loss}
\end{figure*}

Figure~\ref{fig:no_o3_loss} shows that the \ce{NO + O3} loss pathway is generally a minor contributor to total atmospheric \ce{O3} loss in most runs with abiotic deposition velocities. Its contribution increases in several of the more oxidized biotic cases, particularly at low and intermediate methane fluxes. This behavior is consistent with the elevated \ce{HNO3} concentrations seen in Figure~\ref{fig:spaghetti} and indicates that \ce{NO_x} chemistry matters most under specific stellar SEDs and oxidizing states. At the highest methane fluxes, however, little \ce{O3} remains because strong \ce{CH4} input suppresses oxidant buildup, limiting the importance of the \ce{NO + O3} sink.

\section{FUV/NUV Trends in Atmospheric Column Densities}
\label{app:fuvnuv_columns}

\begin{figure*}[t!]
    \centering
    \includegraphics[width=\linewidth]{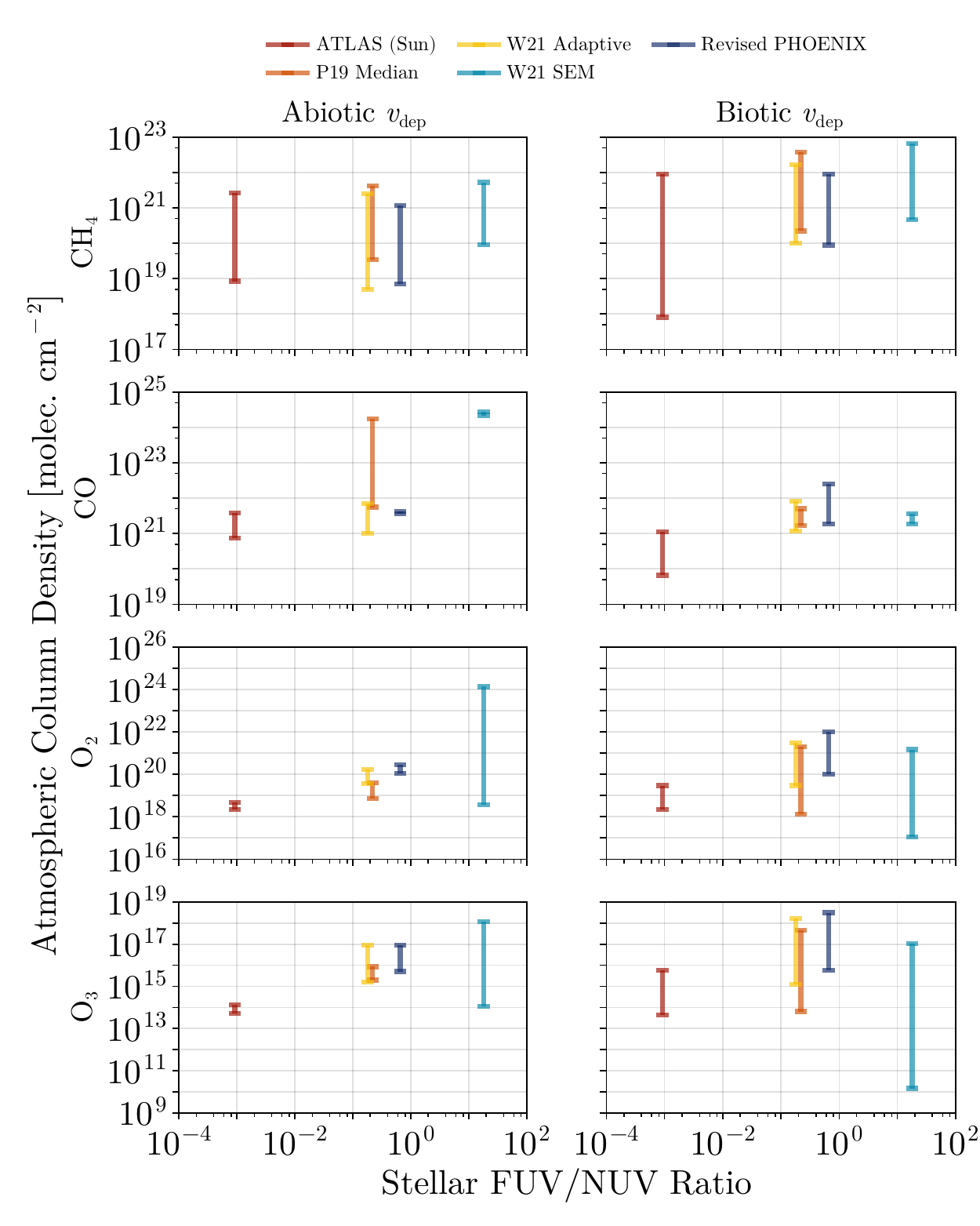}
    \caption{Atmospheric column densities of \ce{CH4}, \ce{CO}, \ce{O2}, and \ce{O3} are plotted as a function of stellar FUV/NUV ratio for the same photochemical model suite shown in Figure~\ref{fig:fuvnuv}. The far-ultraviolet is defined from \num{120.0} to \qty{175.0}{\nano\meter}, and the near-ultraviolet is defined from \num{175.0} to \qty{312.5}{\nano\meter}. Whiskers span $F_{\ce{CH4}}=\num{1E8}$--\qty{1E10}{\molecule\per\centi\meter\squared\per\second} for cases with abiotic deposition velocities (left) and $F_{\ce{CH4}}=\num{1E9}$--\qty{1E11}{\molecule\per\centi\meter\squared\per\second} for cases with biotic deposition velocities (right). The column-depth trends broadly mirror the surface mixing ratio trends in Figure~\ref{fig:fuvnuv}.}
    \label{fig:fuvnuv_column_densities}
\end{figure*}

Figure~\ref{fig:fuvnuv_column_densities} shows the atmospheric column densities of \ce{CH4}, \ce{CO}, \ce{O2}, and \ce{O3} as a function of the stellar FUV/NUV ratio, using the same model suite and plotting convention as Figure~\ref{fig:fuvnuv}.
This comparison tests whether the trends inferred from surface volume mixing ratios persist when the full vertical atmospheric column is considered.
The broad behavior is similar to the surface-abundance comparison: high-FUV/NUV cases with abiotic deposition velocities produce among the largest \ce{O2} and \ce{O3} columns, while cases with biotic deposition velocities show much smaller \ce{CO} columns because of enhanced surface uptake. The biotic cases also produce large \ce{O3} column ranges for several SEDs, underscoring that column-integrated oxidant signatures depend on both the stellar spectrum and lower-boundary redox conditions.
The column-depth view is especially relevant for interpreting the synthetic transmission spectra, since spectral feature strengths depend on the integrated abundance and vertical distribution of absorbers rather than on surface mixing ratios alone.

\section{Atmospheric Conditions}

\begin{deluxetable*}{cccc}[t!]
\tablecaption{Atmospheric Boundary Conditions}
\tablehead{\colhead{Species} & \colhead{Dry Deposition Velocity} & \colhead{Surface Mixing Ratio} & \colhead{Surface Flux} \\ 
\colhead{} & \colhead{$\left[\textrm{cm s}^{-1}\right]$} & \colhead{} & \colhead{$\left[\textrm{molecules cm}^{-2} \textrm{ s}^{-1}\right]$} }  
\label{tab:conditions}
\startdata
\ce{O} & \num{1.0E+00} & - & - \\
\ce{O2} & variable & - & - \\
\ce{H} & \num{1.0E+00} & - & - \\
\ce{OH} & \num{1.0E+00} & - & - \\
\ce{HO2} & \num{1.0E+00} & - & - \\
\ce{H2O2} & \num{2.0E-01} & - & - \\
\ce{H2} & variable & - & \num{1.0E+10} \\
\ce{CO} & variable & - & - \\
\ce{HCO} & \num{1.0E+00} & - & - \\
\ce{H2CO} & \num{1.0E-01} & - & - \\
\ce{CH4} & - & - & variable \\
\ce{CH3} & \num{1.0E+00} & - & - \\
\ce{NO} & \num{3.0E-04} & - & - \\
\ce{NO2} & \num{3.0E-03} & - & - \\
\ce{HNO} & \num{1.0} & - & - \\
\ce{O3} & \num{7.0E-02} & - & - \\
\ce{HNO3} & \num{2.0E-01} & - & - \\
\ce{HS} & \num{3.0E-03} & - & - \\
\ce{H2S} & \num{2.0E-02} & - & \num{3.5E+08} \\
\ce{HSO} & \num{1.0} & - & - \\
\ce{H2SO4} & \num{2.0E-01} & - & - \\
\ce{SO2} & \num{1.0E+00} & - & \num{3.5E+09} \\
\ce{SO} & \num{3.0E-04} & - & - \\
\ce{CO2} & - & \num{1.0E-01} & - \\
\ce{SO4} Aerosols & \num{1.0E-02} & - & - \\
\ce{S8} Aerosols & \num{1.0E-02} & - & - \\
\ce{C4H2 Aerosols} & \num{1.0E-02} & - & - \\
\ce{C6H4} Aerosols & \num{1.0E-02} & - & - \\
\enddata
\end{deluxetable*}

Table \ref{tab:conditions} summarizes the lower-boundary atmospheric conditions supplied to the \texttt{Atmos} one-dimensional photochemical module. 
We prescribe species-specific dry deposition velocities $v_{\mathrm{dep}}$, surface volume mixing ratios, and upward surface fluxes $F$. 
All unlisted species have the three quantities set to zero at the lower boundary. 
For the cases shown in Figs.~\ref{fig:spaghetti}, \ref{fig:abiotic_transmission}, \ref{fig:biotic_transmission}, and \ref{fig:fuvnuv}, we adopt $v_{\mathrm{dep}}(\ce{CO})=\qty{1.2e-4}{\centi\meter\per\second}$,  $v_{\mathrm{dep}}(\ce{H2})=\qty{2.4e-4}{\centi\meter\per\second}$, and $v_{\mathrm{dep}}(\ce{O2})=\qty{1.4E-4}{\centi\meter\per\second}$ for the biotic scenarios and $v_{\mathrm{dep}}(\ce{CO})=\qty{1.0e-8}{\centi\meter\per\second}$, $v_{\mathrm{dep}}(\ce{H2})=\qty{0}{\centi\meter\per\second}$, and $v_{\mathrm{dep}}(\ce{O2})=\qty{0}{\centi\meter\per\second}$ for the scenarios with abiotic deposition velocities. 
In these same figures $F_{\ce{CH4}}$ is equal to \numlist{1e8}, \num{1e9}, or \qty{1e10}{\molecule\per\centi\meter\squared\per\second}, when paired with abiotic deposition velocities. $F_{\ce{CH4}}$ is equal to \numlist{1e9}, \num{1e10}, or \qty{1e11}{\molecule\per\centi\meter\squared\per\second}, when paired with biotic deposition velocities except for a single Mega-MUSCLES Adaptive case, where the nominal \num{1e10} run did not converge. For that case, the nearest converged solution, using a \ce{CH4} flux of \qty{1.5e10}{\molecule\per\centi\meter\squared\per\second} is shown instead. 

\clearpage

\bibliography{paper}{}

\begin{thebibliography}{}
\expandafter\ifx\csname natexlab\endcsname\relax\def\natexlab#1{#1}\fi
\providecommand{\url}[1]{\href{#1}{#1}}
\providecommand{\dodoi}[1]{doi:~\href{http://doi.org/#1}{\nolinkurl{#1}}}
\providecommand{\doeprint}[1]{\href{http://ascl.net/#1}{\nolinkurl{http://ascl.net/#1}}}
\providecommand{\doarXiv}[1]{\href{https://arxiv.org/abs/#1}{\nolinkurl{https://arxiv.org/abs/#1}}}

\bibitem[{E. Agol {et~al.}(2021)Agol, Dorn, Grimm, Turbet, Ducrot, Delrez, Gillon, Demory, Burdanov, Barkaoui, Benkhaldoun, Bolmont, Burgasser, Carey, de~Wit, Fabrycky, {Foreman-Mackey}, Haldemann, Hernandez, Ingalls, Jehin, Langford, Leconte, Lederer, Luger, Malhotra, Meadows, Morris, Pozuelos, Queloz, Raymond, Selsis, Sestovic, Triaud, \& Grootel}]{agolRefiningTransittimingPhotometric2021}
Agol, E., Dorn, C., Grimm, S.~L., {et~al.} 2021, \bibinfo{title}{Refining the {{Transit-timing}} and {{Photometric Analysis}} of {{TRAPPIST-1}}: {{Masses}}, {{Radii}}, {{Densities}}, {{Dynamics}}, and {{Ephemerides}},} The Planetary Science Journal, 2, 1, \dodoi{10.3847/PSJ/abd022}

\bibitem[{A.~C. Allwood {et~al.}(2006)Allwood, Walter, Kamber, Marshall, \& Burch}]{allwoodStromatoliteReefEarly2006}
Allwood, A.~C., Walter, M.~R., Kamber, B.~S., Marshall, C.~P., \& Burch, I.~W. 2006, \bibinfo{title}{Stromatolite Reef from the {{Early Archaean}} Era of {{Australia}},} Nature, 441, 714, \dodoi{10.1038/nature04764}

\bibitem[{G. Arney {et~al.}(2016)Arney, {Domagal-Goldman}, Meadows, Wolf, Schwieterman, Charnay, Claire, H{\'e}brard, \& Trainer}]{arneyPaleOrangeDot2016}
Arney, G., {Domagal-Goldman}, S.~D., Meadows, V.~S., {et~al.} 2016, \bibinfo{title}{The {{Pale Orange Dot}}: {{The Spectrum}} and {{Habitability}} of {{Hazy Archean Earth}},} Astrobiology, 16, 873, \dodoi{10.1089/ast.2015.1422}

\bibitem[{G.~N. Arney {et~al.}(2017)Arney, Meadows, {Domagal-Goldman}, Deming, Robinson, Tovar, Wolf, \& Schwieterman}]{arneyPaleOrangeDots2017}
Arney, G.~N., Meadows, V.~S., {Domagal-Goldman}, S.~D., {et~al.} 2017, \bibinfo{title}{Pale {{Orange Dots}}: {{The Impact}} of {{Organic Haze}} on the {{Habitability}} and {{Detectability}} of {{Earthlike Exoplanets}},} The Astrophysical Journal, 836, 49, \dodoi{10.3847/1538-4357/836/1/49}

\bibitem[{P. {Auclair-Desrotour} \& K. Heng(2020){Auclair-Desrotour} \& Heng}]{auclair-desrotourAtmosphericStabilityCollapse2020}
{Auclair-Desrotour}, P., \& Heng, K. 2020, \bibinfo{title}{Atmospheric Stability and Collapse on Tidally Locked Rocky Planets,} Astronomy \& Astrophysics, 638, A77, \dodoi{10.1051/0004-6361/202037513}

\bibitem[{P. Barth {et~al.}(2024)Barth, St{\"u}eken, Helling, Schwieterman, \& Telling}]{barthEffectLightningAtmospheric2024}
Barth, P., St{\"u}eken, E.~E., Helling, C., Schwieterman, E.~W., \& Telling, J. 2024, \bibinfo{title}{The Effect of Lightning on the Atmospheric Chemistry of Exoplanets and Potential Biosignatures,} Astronomy \& Astrophysics, 686, A58, \dodoi{10.1051/0004-6361/202347286}

\bibitem[{C. Battersby {et~al.}(2018)Battersby, Armus, Bergin, Kataria, Meixner, Pope, Stevenson, Cooray, Leisawitz, Scott, Bauer, Bradford, Ennico, Fortney, Kaltenegger, Melnick, Milam, Narayanan, Padgett, Pontoppidan, Roellig, Sandstrom, Su, Vieira, Wright, Zmuidzinas, Staguhn, Sheth, Benford, Mamajek, Neff, Carey, Burgarella, De~Beck, Gerin, Helmich, Moseley, Sakon, \& Wiedner}]{battersbyOriginsSpaceTelescope2018}
Battersby, C., Armus, L., Bergin, E., {et~al.} 2018, \bibinfo{title}{The {{Origins Space Telescope}},} Nature Astronomy, 2, 596, \dodoi{10.1038/s41550-018-0540-y}

\bibitem[{J. Bezanson {et~al.}(2017)Bezanson, Edelman, Karpinski, \& Shah}]{bezansonJuliaFreshApproach2017}
Bezanson, J., Edelman, A., Karpinski, S., \& Shah, V.~B. 2017, \bibinfo{title}{Julia: {{A Fresh Approach}} to {{Numerical Computing}},} SIAM Review, 59, 65, \dodoi{10.1137/141000671}

\bibitem[{V. Bourrier {et~al.}(2017)Bourrier, Ehrenreich, Wheatley, Bolmont, Gillon, de~Wit, Burgasser, Jehin, Queloz, \& Triaud}]{bourrierReconnaissanceTRAPPIST1Exoplanet2017}
Bourrier, V., Ehrenreich, D., Wheatley, P.~J., {et~al.} 2017, \bibinfo{title}{Reconnaissance of the {{TRAPPIST-1}} Exoplanet System in the {{Lyman-$\alpha$}} Line,} Astronomy \& Astrophysics, 599, L3, \dodoi{10.1051/0004-6361/201630238}

\bibitem[{W. Broussard {et~al.}(2024)Broussard, Schwieterman, Ranjan, {Sousa-Silva}, Fateev, \& Reinhard}]{broussardImpactExtendedH2O2024}
Broussard, W., Schwieterman, E.~W., Ranjan, S., {et~al.} 2024, \bibinfo{title}{The {{Impact}} of {{Extended H2O Cross Sections}} on {{Temperate Anoxic Planet Atmospheres}}: {{Implications}} for {{Spectral Characterization}} of {{Habitable Worlds}},} The Astrophysical Journal, 967, 114, \dodoi{10.3847/1538-4357/ad3a65}

\bibitem[{W. Broussard {et~al.}(2025)Broussard, Schwieterman, {Sousa-Silva}, {Sanger-Johnson}, Ranjan, \& Venot}]{broussardImpactExtendedCO22025}
Broussard, W., Schwieterman, E.~W., {Sousa-Silva}, C., {et~al.} 2025, \bibinfo{title}{The {{Impact}} of {{Extended CO2 Cross Sections}} on {{Temperate Anoxic Planet Atmospheres}},} The Astrophysical Journal, 980, 198, \dodoi{10.3847/1538-4357/adaaf0}

\bibitem[{A.~C. Carnall(2017)Carnall}]{carnallSpectResFastSpectral2017}
Carnall, A.~C. 2017, {{SpectRes}}: {{A Fast Spectral Resampling Tool}} in {{Python}}, arXiv, \dodoi{10.48550/arXiv.1705.05165}

\bibitem[{D.~C. Catling \& J.~F. Kasting(2017{\natexlab{a}})Catling \& Kasting}]{catlingPrebioticEarlyPostbiotic2017}
Catling, D.~C., \& Kasting, J.~F. 2017{\natexlab{a}}, \bibinfo{title}{The {{Prebiotic}} and {{Early Postbiotic Atmosphere}},} in Atmospheric {{Evolution}} on {{Inhabited}} and {{Lifeless Worlds}} (Cambridge: Cambridge University Press), 231--256, \dodoi{10.1017/9781139020558.010}

\bibitem[{D.~C. Catling \& J.~F. Kasting(2017{\natexlab{b}})Catling \& Kasting}]{catlingRiseOxygenOzone2017}
Catling, D.~C., \& Kasting, J.~F. 2017{\natexlab{b}}, \bibinfo{title}{The {{Rise}} of {{Oxygen}} and {{Ozone}} in {{Earth}}'s {{Atmosphere}},} in Atmospheric {{Evolution}} on {{Inhabited}} and {{Lifeless Worlds}} (Cambridge: Cambridge University Press), 257--298, \dodoi{10.1017/9781139020558.011}

\bibitem[{D.~C. Catling \& K.~J. Zahnle(2020)Catling \& Zahnle}]{catlingArcheanAtmosphere2020}
Catling, D.~C., \& Zahnle, K.~J. 2020, \bibinfo{title}{The {{Archean}} Atmosphere,} Science Advances, 6, eaax1420, \dodoi{10.1126/sciadv.aax1420}

\bibitem[{B. Cavalazzi {et~al.}(2021)Cavalazzi, Lemelle, Simionovici, Cady, Russell, Bailo, Canteri, Enrico, Manceau, Maris, Salom{\'e}, Thomassot, Bouden, Tucoulou, \& Hofmann}]{cavalazziCellularRemains342billionyearold2021}
Cavalazzi, B., Lemelle, L., Simionovici, A., {et~al.} 2021, \bibinfo{title}{Cellular Remains in a {\textasciitilde}3.42-Billion-Year-Old Subseafloor Hydrothermal Environment,} Science Advances, 7, eabf3963, \dodoi{10.1126/sciadv.abf3963}

\bibitem[{D. Chen \& D.~C. Catling(2026)Chen \& Catling}]{chenNewEmpiricalKinetics2026}
Chen, D., \& Catling, D.~C. 2026, \bibinfo{title}{New Empirical Kinetics of Iron Oxidation by {{CO2}} Applied to Micrometeorites Imply a {{CO2-rich Archean}} Atmosphere,} Geochimica et Cosmochimica Acta, 416, 183, \dodoi{10.1016/j.gca.2026.01.020}

\bibitem[{T.~A. Collaboration {et~al.}(2018)Collaboration, {Price-Whelan}, Sip{\H o}cz, G{\"u}nther, Lim, Crawford, Conseil, Shupe, Craig, Dencheva, Ginsburg, VanderPlas, Bradley, {P{\'e}rez-Su{\'a}rez}, de~{Val-Borro}, Contributors), Aldcroft, Cruz, Robitaille, Tollerud, Committee), Ardelean, Babej, Bach, Bachetti, Bakanov, Bamford, Barentsen, Barmby, Baumbach, Berry, Biscani, Boquien, Bostroem, Bouma, Brammer, Bray, Breytenbach, Buddelmeijer, Burke, Calderone, Rodr{\'i}guez, Cara, Cardoso, Cheedella, Copin, Corrales, Crichton, D'Avella, Deil, Depagne, Dietrich, Donath, Droettboom, Earl, Erben, Fabbro, Ferreira, Finethy, Fox, Garrison, Gibbons, Goldstein, Gommers, Greco, Greenfield, Groener, Grollier, Hagen, Hirst, Homeier, Horton, Hosseinzadeh, Hu, Hunkeler, Ivezi{\'c}, Jain, Jenness, Kanarek, Kendrew, Kern, Kerzendorf, Khvalko, King, Kirkby, Kulkarni, Kumar, Lee, Lenz, Littlefair, Ma, Macleod, Mastropietro, McCully, Montagnac, Morris, Mueller, Mumford, Muna, Murphy, Nelson, Nguyen, Ninan, N{\"o}the, Ogaz,
  Oh, Parejko, Parley, Pascual, Patil, Patil, Plunkett, Prochaska, Rastogi, Janga, Sabater, Sakurikar, Seifert, Sherbert, {Sherwood-Taylor}, Shih, Sick, Silbiger, Singanamalla, Singer, Sladen, Sooley, Sornarajah, Streicher, Teuben, Thomas, Tremblay, Turner, Terr{\'o}n, van Kerkwijk, de~la Vega, Watkins, Weaver, Whitmore, Woillez, Zabalza, \& Contributors)}]{collaborationAstropyProjectBuilding2018}
Collaboration, T.~A., {Price-Whelan}, A.~M., Sip{\H o}cz, B.~M., {et~al.} 2018, \bibinfo{title}{The {{Astropy Project}}: {{Building}} an {{Open-science Project}} and {{Status}} of the v2.0 {{Core Package}}*,} The Astronomical Journal, 156, 123, \dodoi{10.3847/1538-3881/aabc4f}

\bibitem[{T.~A. Collaboration {et~al.}(2022)Collaboration, {Price-Whelan}, Lim, Earl, Starkman, Bradley, Shupe, Patil, Corrales, Brasseur, N{\"o}the, Donath, Tollerud, Morris, Ginsburg, Vaher, Weaver, Tocknell, Jamieson, van Kerkwijk, Robitaille, Merry, Bachetti, G{\"u}nther, Authors, Aldcroft, {Alvarado-Montes}, Archibald, B{\'o}di, Bapat, Barentsen, Baz{\'a}n, Biswas, Boquien, Burke, Cara, Cara, Conroy, Conseil, Craig, Cross, Cruz, D'Eugenio, Dencheva, Devillepoix, Dietrich, Eigenbrot, Erben, Ferreira, {Foreman-Mackey}, Fox, Freij, Garg, Geda, Glattly, Gondhalekar, Gordon, Grant, Greenfield, Groener, Guest, Gurovich, Handberg, Hart, {Hatfield-Dodds}, Homeier, Hosseinzadeh, Jenness, Jones, Joseph, Kalmbach, Karamehmetoglu, Ka{\l}uszy{\'n}ski, Kelley, Kern, Kerzendorf, Koch, Kulumani, Lee, Ly, Ma, MacBride, Maljaars, Muna, Murphy, Norman, O'Steen, Oman, Pacifici, Pascual, {Pascual-Granado}, Patil, Perren, Pickering, Rastogi, Roulston, Ryan, Rykoff, Sabater, Sakurikar, Salgado, Sanghi, Saunders, Savchenko,
  Schwardt, {Seifert-Eckert}, Shih, Jain, Shukla, Sick, Simpson, Singanamalla, Singer, Singhal, Sinha, Sip{\H o}cz, Spitler, Stansby, Streicher, {\v S}umak, Swinbank, Taranu, Tewary, Tremblay, de~{Val-Borro}, Kooten, Vasovi{\'c}, Verma, Cardoso, Williams, Wilson, Winkel, {Wood-Vasey}, Xue, Yoachim, Zhang, Zonca, \& Contributors}]{collaborationAstropyProjectSustaining2022}
Collaboration, T.~A., {Price-Whelan}, A.~M., Lim, P.~L., {et~al.} 2022, \bibinfo{title}{The {{Astropy Project}}: {{Sustaining}} and {{Growing}} a {{Community-oriented Open-source Project}} and the {{Latest Major Release}} (v5.0) of the {{Core Package}}*,} The Astrophysical Journal, 935, 167, \dodoi{10.3847/1538-4357/ac7c74}

\bibitem[{G.~J. Cooke {et~al.}(2023)Cooke, Marsh, Walsh, \& Youngblood}]{cookeDegenerateInterpretationsO32023}
Cooke, G.~J., Marsh, D.~R., Walsh, C., \& Youngblood, A. 2023, \bibinfo{title}{Degenerate {{Interpretations}} of {{O3 Spectral Features}} in {{Exoplanet Atmosphere Observations Due}} to {{Stellar UV Uncertainties}}: {{A 3D Case Study}} with {{TRAPPIST-1}} e,} The Astrophysical Journal, 959, 45, \dodoi{10.3847/1538-4357/ad0381}

\bibitem[{D. Crisp(1997)Crisp}]{crispAbsorptionSunlightWater1997}
Crisp, D. 1997, \bibinfo{title}{Absorption of Sunlight by Water Vapor in Cloudy Conditions: {{A}} Partial Explanation for the Cloud Absorption Anomaly,} Geophysical Research Letters, 24, 571, \dodoi{10.1029/97GL50245}

\bibitem[{M.~H. Currie {et~al.}(2023)Currie, Meadows, \& Rasmussen}]{currieTheresMoreLife2023}
Currie, M.~H., Meadows, V.~S., \& Rasmussen, K.~C. 2023, \bibinfo{title}{There's {{More}} to {{Life}} than {{O2}}: {{Simulating}} the {{Detectability}} of a {{Range}} of {{Molecules}} for {{Ground-based}}, {{High-resolution Spectroscopy}} of {{Transiting Terrestrial Exoplanets}},} The Planetary Science Journal, 4, 83, \dodoi{10.3847/PSJ/accf86}

\bibitem[{S. Danisch \& J. Krumbiegel(2021)Danisch \& Krumbiegel}]{danischMakieJlFlexible2021}
Danisch, S., \& Krumbiegel, J. 2021, \bibinfo{title}{Makie.Jl: {{Flexible}} High-Performance Data Visualization for {{Julia}},} Journal of Open Source Software, 6, 3349, \dodoi{10.21105/joss.03349}

\bibitem[{J. {de Wit} {et~al.}(2016){de Wit}, Wakeford, Gillon, Lewis, Valenti, Demory, Burgasser, Burdanov, Delrez, Jehin, Lederer, Queloz, Triaud, \& Van~Grootel}]{dewitCombinedTransmissionSpectrum2016}
{de Wit}, J., Wakeford, H.~R., Gillon, M., {et~al.} 2016, \bibinfo{title}{A Combined Transmission Spectrum of the {{Earth-sized}} Exoplanets {{TRAPPIST-1}} b and c,} Nature, 537, 69, \dodoi{10.1038/nature18641}

\bibitem[{J. {de Wit} {et~al.}(2018){de Wit}, Wakeford, Lewis, Delrez, Gillon, Selsis, Leconte, Demory, Bolmont, Bourrier, Burgasser, Grimm, Jehin, Lederer, Owen, Stamenkovi{\'c}, \& Triaud}]{dewitAtmosphericReconnaissanceHabitablezone2018}
{de Wit}, J., Wakeford, H.~R., Lewis, N.~K., {et~al.} 2018, \bibinfo{title}{Atmospheric Reconnaissance of the Habitable-Zone {{Earth-sized}} Planets Orbiting {{TRAPPIST-1}},} Nature Astronomy, 2, 214, \dodoi{10.1038/s41550-017-0374-z}

\bibitem[{L.~N.~R. {do Amaral} {et~al.}(2022){do Amaral}, Barnes, Segura, \& Luger}]{doamaralContributionMdwarfFlares2022}
{do Amaral}, L. N.~R., Barnes, R., Segura, A., \& Luger, R. 2022, \bibinfo{title}{The {{Contribution}} of {{M-dwarf Flares}} to the {{Thermal Escape}} of {{Potentially Habitable Planet Atmospheres}},} The Astrophysical Journal, 928, 12, \dodoi{10.3847/1538-4357/ac53af}

\bibitem[{S.~D. {Domagal-Goldman} {et~al.}(2014){Domagal-Goldman}, Segura, Claire, Robinson, \& Meadows}]{domagal-goldmanAbioticOzoneOxygen2014}
{Domagal-Goldman}, S.~D., Segura, A., Claire, M.~W., Robinson, T.~D., \& Meadows, V.~S. 2014, \bibinfo{title}{Abiotic {{Ozone}} and {{Oxygen}} in {{Atmospheres Similar}} to {{Prebiotic Earth}},} The Astrophysical Journal, 792, 90, \dodoi{10.1088/0004-637X/792/2/90}

\bibitem[{E. Ducrot {et~al.}(2020)Ducrot, Gillon, Delrez, Agol, Rimmer, Turbet, G{\"u}nther, Demory, Triaud, Bolmont, Burgasser, Carey, Ingalls, Jehin, Leconte, Lederer, Queloz, Raymond, Selsis, Grootel, \& de~Wit}]{ducrotTRAPPIST1GlobalResults2020}
Ducrot, E., Gillon, M., Delrez, L., {et~al.} 2020, \bibinfo{title}{{{TRAPPIST-1}}: {{Global}} Results of the {{Spitzer Exploration Science Program Red Worlds}},} Astronomy \& Astrophysics, 640, A112, \dodoi{10.1051/0004-6361/201937392}

\bibitem[{J.~K. {Eager-Nash} {et~al.}(2024){Eager-Nash}, Daines, McDermott, Andrews, Grain, Bishop, Rogers, Smith, Khalek, Boxer, Mak, Ridgway, H{\'e}brard, Lambert, Lenton, \& Mayne}]{eager-nashSimulatingBiosignaturesPreoxygen2024}
{Eager-Nash}, J.~K., Daines, S.~J., McDermott, J.~W., {et~al.} 2024, \bibinfo{title}{Simulating Biosignatures from Pre-Oxygen Photosynthesizing Life on {{TRAPPIST-1e}},} Monthly Notices of the Royal Astronomical Society, 531, 468, \dodoi{10.1093/mnras/stae1142}

\bibitem[{N. Espinoza {et~al.}(2025)Espinoza, Allen, Glidden, Lewis, Seager, Ca{\~n}as, Grant, Gressier, Courreges, Stevenson, Ranjan, Col{\'o}n, Morris, MacDonald, Long, Wakeford, Valenti, Alderson, Batalha, Challener, Huang, Lin, Louie, Mullens, Valentine, Mountain, Pueyo, Perrin, Bellini, Kammerer, Libralato, Rebollido, Rickman, Sohn, \& {van der Marel}}]{espinozaJWSTTSTDREAMSNIRSpec2025}
Espinoza, N., Allen, N.~H., Glidden, A., {et~al.} 2025, \bibinfo{title}{{{JWST-TST DREAMS}}: {{NIRSpec}}/{{PRISM Transmission Spectroscopy}} of the {{Habitable Zone Planet TRAPPIST-1}} e,} The Astrophysical Journal, 990, L52, \dodoi{10.3847/2041-8213/adf42e}

\bibitem[{T.~J. Fauchez {et~al.}(2019)Fauchez, Turbet, Villanueva, Wolf, Arney, Kopparapu, Lincowski, Mandell, de~Wit, Pidhorodetska, {Domagal-Goldman}, \& Stevenson}]{fauchezImpactCloudsHazes2019}
Fauchez, T.~J., Turbet, M., Villanueva, G.~L., {et~al.} 2019, \bibinfo{title}{Impact of {{Clouds}} and {{Hazes}} on the {{Simulated JWST Transmission Spectra}} of {{Habitable Zone Planets}} in the {{TRAPPIST-1 System}},} The Astrophysical Journal, 887, 194, \dodoi{10.3847/1538-4357/ab5862}

\bibitem[{K. France {et~al.}(2016)France, Loyd, Youngblood, Brown, Schneider, Hawley, Froning, Linsky, Roberge, Buccino, Davenport, Fontenla, Kaltenegger, Kowalski, Mauas, Miguel, Redfield, Rugheimer, Tian, Vieytes, Walkowicz, \& Weisenburger}]{franceMUSCLESTREASURYSURVEY2016}
France, K., Loyd, R. O.~P., Youngblood, A., {et~al.} 2016, \bibinfo{title}{{{THE MUSCLES TREASURY SURVEY}}. {{I}}. {{MOTIVATION AND OVERVIEW}}*,} The Astrophysical Journal, 820, 89, \dodoi{10.3847/0004-637X/820/2/89}

\bibitem[{S. Gebauer {et~al.}(2018)Gebauer, Grenfell, Lehmann, \& Rauer}]{gebauerEvolutionEarthlikePlanetary2018}
Gebauer, S., Grenfell, J., Lehmann, R., \& Rauer, H. 2018, \bibinfo{title}{Evolution of {{Earth-like Planetary Atmospheres}} around {{M Dwarf Stars}}: {{Assessing}} the {{Atmospheres}} and {{Biospheres}} with a {{Coupled Atmosphere Biogeochemical Model}},} Astrobiology, 18, 856, \dodoi{10.1089/ast.2017.1723}

\bibitem[{M. Gillon {et~al.}(2016)Gillon, Jehin, Lederer, Delrez, {de Wit}, Burdanov, Van~Grootel, Burgasser, Triaud, Opitom, Demory, Sahu, Bardalez~Gagliuffi, Magain, \& Queloz}]{gillonTemperateEarthsizedPlanets2016}
Gillon, M., Jehin, E., Lederer, S.~M., {et~al.} 2016, \bibinfo{title}{Temperate {{Earth-sized}} Planets Transiting a Nearby Ultracool Dwarf Star,} Nature, 533, 221, \dodoi{10.1038/nature17448}

\bibitem[{M. Gillon {et~al.}(2017)Gillon, Triaud, Demory, Jehin, Agol, Deck, Lederer, de~Wit, Burdanov, Ingalls, Bolmont, Leconte, Raymond, Selsis, Turbet, Barkaoui, Burgasser, Burleigh, Carey, Chaushev, Copperwheat, Delrez, Fernandes, Holdsworth, Kotze, Grootel, Almleaky, Benkhaldoun, Magain, \& Queloz}]{gillonSevenTemperateTerrestrial2017}
Gillon, M., Triaud, A. H. M.~J., Demory, B.-O., {et~al.} 2017, \bibinfo{title}{Seven Temperate Terrestrial Planets around the Nearby Ultracool Dwarf Star {{TRAPPIST-1}},} Nature, 542, 456, \dodoi{10.1038/nature21360}

\bibitem[{M. Gillon {et~al.}(2026)Gillon, Ducrot, Bell, Huang, Lincowski, Lyu, Maurel, Revol, Agol, Bolmont, Dong, Fauchez, Koll, Leconte, Meadows, Selsis, Turbet, Charnay, Delrez, Demory, Householder, Zieba, Berardo, Dyrek, Edwards, {de Wit}, Greene, Hu, Iro, Kreidberg, Lagage, {Lustig-Yaeger}, \& Iyer}]{gillonNoThickAtmosphere2026}
Gillon, M., Ducrot, E., Bell, T.~J., {et~al.} 2026, \bibinfo{title}{No Thick Atmosphere around {{TRAPPIST-1}} b and c from {{JWST}} Thermal Phase Curves,} Nature Astronomy, 1, \dodoi{10.1038/s41550-026-02806-9}

\bibitem[{A. Glidden {et~al.}(2025)Glidden, Ranjan, Seager, Espinoza, MacDonald, Allen, Ca{\~n}as, Grant, Gressier, Stevenson, Batalha, Lewis, Long, Wakeford, Alderson, Challener, Col{\'o}n, Huang, Lin, Louie, Mullens, Sotzen, Valenti, Valentine, Clampin, Mountain, Perrin, \& {van der Marel}}]{gliddenJWSTTSTDREAMSSecondary2025}
Glidden, A., Ranjan, S., Seager, S., {et~al.} 2025, \bibinfo{title}{{{JWST-TST DREAMS}}: {{Secondary Atmosphere Constraints}} for the {{Habitable Zone Planet TRAPPIST-1}} e,} The Astrophysical Journal Letters, 990, L53, \dodoi{10.3847/2041-8213/adf62e}

\bibitem[{I.~E. Gordon {et~al.}(2022)Gordon, Rothman, Hargreaves, Hashemi, Karlovets, Skinner, Conway, Hill, Kochanov, Tan, Wcis{\l}o, Finenko, Nelson, Bernath, Birk, Boudon, Campargue, Chance, Coustenis, Drouin, Flaud, Gamache, Hodges, Jacquemart, Mlawer, Nikitin, Perevalov, Rotger, Tennyson, Toon, Tran, Tyuterev, Adkins, Baker, Barbe, Can{\`e}, Cs{\'a}sz{\'a}r, Dudaryonok, Egorov, Fleisher, Fleurbaey, Foltynowicz, Furtenbacher, Harrison, Hartmann, Horneman, Huang, Karman, Karns, Kassi, Kleiner, Kofman, {Kwabia--Tchana}, Lavrentieva, Lee, Long, Lukashevskaya, Lyulin, Makhnev, Matt, Massie, Melosso, Mikhailenko, Mondelain, M{\"u}ller, Naumenko, Perrin, Polyansky, Raddaoui, Raston, Reed, Rey, Richard, T{\'o}bi{\'a}s, Sadiek, Schwenke, Starikova, Sung, Tamassia, Tashkun, Vander~Auwera, Vasilenko, Vigasin, Villanueva, Vispoel, Wagner, Yachmenev, \& Yurchenko}]{gordonHITRAN2020MolecularSpectroscopic2022}
Gordon, I.~E., Rothman, L.~S., Hargreaves, R.~J., {et~al.} 2022, \bibinfo{title}{The {{HITRAN2020}} Molecular Spectroscopic Database,} Journal of Quantitative Spectroscopy and Radiative Transfer, 277, 107949, \dodoi{10.1016/j.jqsrt.2021.107949}

\bibitem[{T.~P. Greene {et~al.}(2023)Greene, Bell, Ducrot, Dyrek, Lagage, \& Fortney}]{greeneThermalEmissionEarthsized2023}
Greene, T.~P., Bell, T.~J., Ducrot, E., {et~al.} 2023, \bibinfo{title}{Thermal Emission from the {{Earth-sized}} Exoplanet {{TRAPPIST-1}} b Using {{JWST}},} Nature, 618, 39, \dodoi{10.1038/s41586-023-05951-7}

\bibitem[{J.~L. Grenfell {et~al.}(2014)Grenfell, Gebauer, {v. Paris}, Godolt, \& Rauer}]{grenfellSensitivityBiosignaturesEarthlike2014}
Grenfell, J.~L., Gebauer, S., {v. Paris}, P., Godolt, M., \& Rauer, H. 2014, \bibinfo{title}{Sensitivity of Biosignatures on {{Earth-like}} Planets Orbiting in the Habitable Zone of Cool {{M-dwarf Stars}} to Varying Stellar {{UV}} Radiation and Surface Biomass Emissions,} Planetary and Space Science, 98, 66, \dodoi{10.1016/j.pss.2013.10.006}

\bibitem[{K.~K. {Hardegree-Ullman} {et~al.}(2023){Hardegree-Ullman}, Apai, Bergsten, Pascucci, \& {L{\'o}pez-Morales}}]{hardegree-ullmanBioverseComprehensiveAssessment2023}
{Hardegree-Ullman}, K.~K., Apai, D., Bergsten, G.~J., Pascucci, I., \& {L{\'o}pez-Morales}, M. 2023, \bibinfo{title}{Bioverse: {{A Comprehensive Assessment}} of the {{Capabilities}} of {{Extremely Large Telescopes}} to {{Probe Earth-like O2 Levels}} in {{Nearby Transiting Habitable-zone Exoplanets}},} The Astronomical Journal, 165, 267, \dodoi{10.3847/1538-3881/acd1ec}

\bibitem[{C.~E. Harman {et~al.}(2018)Harman, Felton, Hu, {Domagal-Goldman}, Segura, Tian, \& Kasting}]{harmanAbioticO2Levels2018}
Harman, C.~E., Felton, R., Hu, R., {et~al.} 2018, \bibinfo{title}{Abiotic {{O2 Levels}} on {{Planets}} around {{F}}, {{G}}, {{K}}, and {{M Stars}}: {{Effects}} of {{Lightning-produced Catalysts}} in {{Eliminating Oxygen False Positives}},} The Astrophysical Journal, 866, 56, \dodoi{10.3847/1538-4357/aadd9b}

\bibitem[{C.~E. Harman {et~al.}(2015)Harman, Schwieterman, Schottelkotte, \& Kasting}]{harmanAbioticO2Levels2015}
Harman, C.~E., Schwieterman, E.~W., Schottelkotte, J.~C., \& Kasting, J.~F. 2015, \bibinfo{title}{Abiotic {{O2 Levels}} on {{Planets}} around {{F}}, {{G}}, {{K}}, and {{M Stars}}: {{Possible False Positives}} for {{Life}}?} The Astrophysical Journal, 812, 137, \dodoi{10.1088/0004-637X/812/2/137}

\bibitem[{C.~R. Harris {et~al.}(2020)Harris, Millman, {van der Walt}, Gommers, Virtanen, Cournapeau, Wieser, Taylor, Berg, Smith, Kern, Picus, Hoyer, {van Kerkwijk}, Brett, Haldane, {del R{\'i}o}, Wiebe, Peterson, {G{\'e}rard-Marchant}, Sheppard, Reddy, Weckesser, Abbasi, Gohlke, \& Oliphant}]{harrisArrayProgrammingNumPy2020}
Harris, C.~R., Millman, K.~J., {van der Walt}, S.~J., {et~al.} 2020, \bibinfo{title}{Array Programming with {{NumPy}},} Nature, 585, 357, \dodoi{10.1038/s41586-020-2649-2}

\bibitem[{C.~K. Jones {et~al.}(2025)Jones, Leung, Ebadirad, Sneed, \& Lyons}]{jonesEvolutionEarthsAtmosphere2025}
Jones, C.~K., Leung, M., Ebadirad, S., Sneed, E.~L., \& Lyons, T.~W. 2025, \bibinfo{title}{Evolution of {{Earth}}'s {{Atmosphere}},} in Encyclopedia of {{Atmospheric Sciences}}, Third edition edn., ed. W.~A. Robinson \& P.~Yang (Academic Press), 790--803, \dodoi{10.1016/B978-0-323-96026-7.00226-5}

\bibitem[{J.~F. Kasting {et~al.}(1979)Kasting, Liu, \& Donahue}]{kastingOxygenLevelsPrebiological1979}
Kasting, J.~F., Liu, S.~C., \& Donahue, T.~M. 1979, \bibinfo{title}{Oxygen Levels in the Prebiological Atmosphere,} Journal of Geophysical Research: Oceans, 84, 3097, \dodoi{10.1029/JC084iC06p03097}

\bibitem[{J.~F. Kasting {et~al.}(1983)Kasting, Zahnle, \& Walker}]{kastingPhotochemistryMethaneEarths1983}
Kasting, J.~F., Zahnle, K.~J., \& Walker, J. C.~G. 1983, \bibinfo{title}{Photochemistry of Methane in the {{Earth}}'s Early Atmosphere,} Precambrian Research, 20, 121, \dodoi{10.1016/0301-9268(83)90069-4}

\bibitem[{P. Kharecha {et~al.}(2005)Kharecha, Kasting, \& Siefert}]{kharechaCoupledAtmosphereEcosystem2005}
Kharecha, P., Kasting, J., \& Siefert, J. 2005, \bibinfo{title}{A Coupled Atmosphere--Ecosystem Model of the Early {{Archean Earth}},} Geobiology, 3, 53, \dodoi{10.1111/j.1472-4669.2005.00049.x}

\bibitem[{R.~K. Kopparapu {et~al.}(2013)Kopparapu, Ramirez, Kasting, Eymet, Robinson, Mahadevan, Terrien, {Domagal-Goldman}, Meadows, \& Deshpande}]{kopparapuHabitableZonesMainSequence2013}
Kopparapu, R.~K., Ramirez, R., Kasting, J.~F., {et~al.} 2013, \bibinfo{title}{Habitable {{Zones}} around {{Main-Sequence Stars}}: {{New Estimates}},} The Astrophysical Journal, 765, 131, \dodoi{10.1088/0004-637X/765/2/131}

\bibitem[{T. Kozakis {et~al.}(2022)Kozakis, Mendon{\c c}a, \& Buchhave}]{kozakisOzoneReliableProxy2022}
Kozakis, T., Mendon{\c c}a, J.~M., \& Buchhave, L.~A. 2022, \bibinfo{title}{Is Ozone a Reliable Proxy for Molecular Oxygen? - {{I}}. {{The O2}}--{{O3}} Relationship for {{Earth-like}} Atmospheres,} Astronomy \& Astrophysics, 665, A156, \dodoi{10.1051/0004-6361/202244164}

\bibitem[{J. {Krissansen-Totton}(2023){Krissansen-Totton}}]{krissansen-tottonImplicationsAtmosphericNondetections2023}
{Krissansen-Totton}, J. 2023, \bibinfo{title}{Implications of {{Atmospheric Nondetections}} for {{Trappist-1 Inner Planets}} on {{Atmospheric Retention Prospects}} for {{Outer Planets}},} The Astrophysical Journal Letters, 951, L39, \dodoi{10.3847/2041-8213/acdc26}

\bibitem[{J. {Krissansen-Totton} {et~al.}(2018{\natexlab{a}}){Krissansen-Totton}, Garland, Irwin, \& Catling}]{krissansen-tottonDetectabilityBiosignaturesAnoxic2018}
{Krissansen-Totton}, J., Garland, R., Irwin, P., \& Catling, D.~C. 2018{\natexlab{a}}, \bibinfo{title}{Detectability of {{Biosignatures}} in {{Anoxic Atmospheres}} with the {{James Webb Space Telescope}}: {{A TRAPPIST-1e Case Study}},} The Astronomical Journal, 156, 114, \dodoi{10.3847/1538-3881/aad564}

\bibitem[{J. {Krissansen-Totton} {et~al.}(2018{\natexlab{b}}){Krissansen-Totton}, Olson, \& Catling}]{krissansen-tottonDisequilibriumBiosignaturesEarth2018}
{Krissansen-Totton}, J., Olson, S., \& Catling, D.~C. 2018{\natexlab{b}}, \bibinfo{title}{Disequilibrium Biosignatures over {{Earth}} History and Implications for Detecting Exoplanet Life,} Science Advances, 4, eaao5747, \dodoi{10.1126/sciadv.aao5747}

\bibitem[{D.~T. Leisawitz {et~al.}(2021)Leisawitz, Amatucci, Allen, Arenberg, Armus, Battersby, Bauer, Beaman, Jr, Beltran, Benford, Bergin, Bolognese, Bradford, Bradley, Burgarella, Carey, Carter, Chi, Cooray, Corsetti, D'Asto, Beck, Denis, Derkacz, Dewell, DiPirro, Earle, East, Edgington, Ennico, Fantano, Feller, Folta, Fortney, Gavares, Generie, Gerin, Granger, Greene, Griffiths, Harpole, Harvey, Helmich, Hilliard, Howard, Jacoby, Jamil, Jamison, Kaltenegger, Kataria, Knight, Knollenberg, Lawrence, Lightsey, Lipscy, Mamajek, Martins, Mather, Meixner, Melnick, Milam, Mooney, Moseley, Narayanan, Neff, Nguyen, Nordt, Olson, Padgett, Petach, Petro, Pohner, Pontoppidan, Pope, Ramspacker, Rao, Roellig, Sakon, Sandin, Sandstrom, Scott, Seals, Sheth, Sokolsky, Staguhn, Steeves, Stevenson, Stoneking, Su, Tajdaran, Tompkins, Vieira, Webster, Wiedner, Wright, Wu, \& Zmuidzinas}]{leisawitzOriginsSpaceTelescope2021}
Leisawitz, D.~T., Amatucci, E.~G., Allen, L.~N., {et~al.} 2021, \bibinfo{title}{Origins {{Space Telescope}}: Baseline Mission Concept,} Journal of Astronomical Telescopes, Instruments, and Systems, 7, 011002, \dodoi{10.1117/1.JATIS.7.1.011002}

\bibitem[{Z. Lin {et~al.}(2021)Lin, MacDonald, Kaltenegger, \& Wilson}]{linDifferentiatingModernPrebiotic2021a}
Lin, Z., MacDonald, R.~J., Kaltenegger, L., \& Wilson, D.~J. 2021, \bibinfo{title}{Differentiating Modern and Prebiotic {{Earth}} Scenarios for {{TRAPPIST-1e}}: High-Resolution Transmission Spectra and Predictions for {{JWST}},} Monthly Notices of the Royal Astronomical Society, 505, 3562, \dodoi{10.1093/mnras/stab1486}

\bibitem[{A.~P. Lincowski {et~al.}(2018)Lincowski, Meadows, Crisp, Robinson, Luger, {Lustig-Yaeger}, \& Arney}]{lincowskiEvolvedClimatesObservational2018}
Lincowski, A.~P., Meadows, V.~S., Crisp, D., {et~al.} 2018, \bibinfo{title}{Evolved {{Climates}} and {{Observational Discriminants}} for the {{TRAPPIST-1 Planetary System}},} The Astrophysical Journal, 867, 76, \dodoi{10.3847/1538-4357/aae36a}

\bibitem[{T.~W. Lyons {et~al.}(2014)Lyons, Reinhard, \& Planavsky}]{lyonsRiseOxygenEarth2014}
Lyons, T.~W., Reinhard, C.~T., \& Planavsky, N.~J. 2014, \bibinfo{title}{The Rise of Oxygen in {{Earth}}'s Early Ocean and Atmosphere,} Nature, 506, 307, \dodoi{10.1038/nature13068}

\bibitem[{M.~T. Mak {et~al.}(2024)Mak, Sergeev, Mayne, Banks, {Eager-Nash}, Manners, Arney, H{\'e}brard, \& Kohary}]{mak3DSimulationsTRAPPIST1e2024}
Mak, M.~T., Sergeev, D.~E., Mayne, N., {et~al.} 2024, \bibinfo{title}{{{3D}} Simulations of {{TRAPPIST-1e}} with Varying {{CO2}}, {{CH4}}, and Haze Profiles,} Monthly Notices of the Royal Astronomical Society, 529, 3971, \dodoi{10.1093/mnras/stae741}

\bibitem[{V.~S. Meadows(2017)Meadows}]{meadowsReflectionsO2Biosignature2017}
Meadows, V.~S. 2017, \bibinfo{title}{Reflections on {{O2}} as a {{Biosignature}} in {{Exoplanetary Atmospheres}},} Astrobiology, 17, 1022, \dodoi{10.1089/ast.2016.1578}

\bibitem[{V.~S. Meadows \& D. Crisp(1996)Meadows \& Crisp}]{meadowsGroundbasedNearinfraredObservations1996}
Meadows, V.~S., \& Crisp, D. 1996, \bibinfo{title}{Ground-Based near-Infrared Observations of the {{Venus}} Nightside: {{The}} Thermal Structure and Water Abundance near the Surface,} Journal of Geophysical Research: Planets, 101, 4595, \dodoi{10.1029/95JE03567}

\bibitem[{V.~S. Meadows {et~al.}(2023)Meadows, Lincowski, \& {Lustig-Yaeger}}]{meadowsFeasibilityDetectingBiosignatures2023}
Meadows, V.~S., Lincowski, A.~P., \& {Lustig-Yaeger}, J. 2023, \bibinfo{title}{The {{Feasibility}} of {{Detecting Biosignatures}} in the {{TRAPPIST-1 Planetary System}} with {{JWST}},} The Planetary Science Journal, 4, 192, \dodoi{10.3847/PSJ/acf488}

\bibitem[{V.~S. Meadows {et~al.}(2018)Meadows, Reinhard, Arney, Parenteau, Schwieterman, {Domagal-Goldman}, Lincowski, Stapelfeldt, Rauer, DasSarma, Hegde, Narita, Deitrick, {Lustig-Yaeger}, Lyons, Siegler, \& Grenfell}]{meadowsExoplanetBiosignaturesUnderstanding2018}
Meadows, V.~S., Reinhard, C.~T., Arney, G.~N., {et~al.} 2018, \bibinfo{title}{Exoplanet {{Biosignatures}}: {{Understanding Oxygen}} as a {{Biosignature}} in the {{Context}} of {{Its Environment}},} Astrobiology, 18, 630, \dodoi{10.1089/ast.2017.1727}

\bibitem[{M. Meixner {et~al.}(2019)Meixner, Cooray, Leisawitz, Staguhn, Armus, Battersby, Bauer, Bergin, Bradford, {Ennico-Smith}, Fortney, Kataria, Melnick, Milam, Narayanan, Padgett, Pontoppidan, Pope, Roellig, Sandstrom, Stevenson, Su, Vieira, Wright, Zmuidzinas, Sheth, Benford, Mamajek, Neff, De~Beck, Gerin, Helmich, Sakon, Scott, Vavrek, Wiedner, Carey, Burgarella, Moseley, Amatucci, Carter, DiPirro, Wu, Beaman, Beltran, Bolognese, Bradley, Corsetti, D'Asto, Denis, Derkacz, Earle, Fantano, Folta, Gavares, Generie, Hilliard, Howard, Jamil, Jamison, Lynch, Martins, Petro, Ramspacher, Rao, Sandin, Stoneking, Tompkins, \& Webster}]{meixnerOriginsSpaceTelescope2019}
Meixner, M., Cooray, A., Leisawitz, D., {et~al.} 2019, Origins {{Space Telescope Mission Concept Study Report}},, https://arxiv.org/abs/1912.06213v2

\bibitem[{A. {Miranda-Rosete} {et~al.}(2025){Miranda-Rosete}, Segura, \& Schwieterman}]{miranda-roseteBiosignatureFalsePositives2025}
{Miranda-Rosete}, A., Segura, A., \& Schwieterman, E.~W. 2025, \bibinfo{title}{Biosignature {{False Positives}} in {{Potentially Habitable Planets}} around {{M Dwarfs}}: {{The Effect}} of {{UV Radiation}} from {{One Flare}},} The Astrophysical Journal, 989, 34, \dodoi{10.3847/1538-4357/acebec}

\bibitem[{V. Molina {et~al.}(2026)Molina, Aguilar, Dorador, {Tregloan-Reed}, {Rojas-Ayala}, Carcamo, Aliaga, \& Soto}]{molinaExploringExtremophileGas2026}
Molina, V., Aguilar, P., Dorador, C., {et~al.} 2026, \bibinfo{title}{Exploring Extremophile Gas Production as a Biomarker for Early {{Earth}} Atmospheres,} International Journal of Astrobiology, 25, e4, \dodoi{10.1017/S1473550425100268}

\bibitem[{N. Noffke {et~al.}(2013)Noffke, Christian, Wacey, \& Hazen}]{noffkeMicrobiallyInducedSedimentary2013}
Noffke, N., Christian, D., Wacey, D., \& Hazen, R.~M. 2013, \bibinfo{title}{Microbially {{Induced Sedimentary Structures Recording}} an {{Ancient Ecosystem}} in the ca. 3.48 {{Billion-Year-Old Dresser Formation}}, {{Pilbara}}, {{Western Australia}},} Astrobiology, 13, 1103, \dodoi{10.1089/ast.2013.1030}

\bibitem[{J.~T. {O'Malley-James} \& L. Kaltenegger(2017){O'Malley-James} \& Kaltenegger}]{omalley-jamesUVSurfaceHabitability2017}
{O'Malley-James}, J.~T., \& Kaltenegger, L. 2017, \bibinfo{title}{{{UV}} Surface Habitability of the {{TRAPPIST-1}} System,} Monthly Notices of the Royal Astronomical Society: Letters, 469, L26, \dodoi{10.1093/mnrasl/slx047}

\bibitem[{E.~K. Pass {et~al.}(2025)Pass, Charbonneau, \& Vanderburg}]{passRecedingCosmicShoreline2025}
Pass, E.~K., Charbonneau, D., \& Vanderburg, A. 2025, \bibinfo{title}{The {{Receding Cosmic Shoreline}} of {{Mid-to-late M Dwarfs}}: {{Measurements}} of {{Active Lifetimes Worsen Challenges}} for {{Atmosphere Retention}} by {{Rocky Exoplanets}},} The Astrophysical Journal Letters, 986, L3, \dodoi{10.3847/2041-8213/adda39}

\bibitem[{A.~A. Pavlov {et~al.}(2001)Pavlov, Brown, \& Kasting}]{pavlovUVShieldingNH32001}
Pavlov, A.~A., Brown, L.~L., \& Kasting, J.~F. 2001, \bibinfo{title}{{{UV}} Shielding of {{NH3}} and {{O2}} by Organic Hazes in the {{Archean}} Atmosphere,} Journal of Geophysical Research: Planets, 106, 23267, \dodoi{10.1029/2000JE001448}

\bibitem[{S. Peacock {et~al.}(2019)Peacock, Barman, Shkolnik, Hauschildt, \& Baron}]{peacockPredictingExtremeUltraviolet2019}
Peacock, S., Barman, T., Shkolnik, E.~L., Hauschildt, P.~H., \& Baron, E. 2019, \bibinfo{title}{Predicting the {{Extreme Ultraviolet Radiation Environment}} of {{Exoplanets}} around {{Low-mass Stars}}: {{The TRAPPIST-1 System}},} The Astrophysical Journal, 871, 235, \dodoi{10.3847/1538-4357/aaf891}

\bibitem[{C. {Piaulet-Ghorayeb} {et~al.}(2025){Piaulet-Ghorayeb}, Benneke, Turbet, Moore, Roy, Lim, Doyon, Fauchez, Albert, Radica, Coulombe, Lafreni{\`e}re, Cowan, Belzile, Musfirat, Kaur, L'Heureux, Johnstone, MacDonald, Allart, Dang, Kaltenegger, Pelletier, Rowe, Taylor, \& Turner}]{piaulet-ghorayebStrictLimitsPotential2025}
{Piaulet-Ghorayeb}, C., Benneke, B., Turbet, M., {et~al.} 2025, \bibinfo{title}{Strict {{Limits}} on {{Potential Secondary Atmospheres}} on the {{Temperate Rocky Exo-Earth TRAPPIST-1}} d,} The Astrophysical Journal, 989, 181, \dodoi{10.3847/1538-4357/adf207}

\bibitem[{S. Ranjan {et~al.}(2020)Ranjan, Schwieterman, Harman, Fateev, {Sousa-Silva}, Seager, \& Hu}]{ranjanPhotochemistryAnoxicAbiotic2020}
Ranjan, S., Schwieterman, E.~W., Harman, C., {et~al.} 2020, \bibinfo{title}{Photochemistry of {{Anoxic Abiotic Habitable Planet Atmospheres}}: {{Impact}} of {{New H2O Cross Sections}},} The Astrophysical Journal, 896, 148, \dodoi{10.3847/1538-4357/ab9363}

\bibitem[{S. Ranjan {et~al.}(2022)Ranjan, Seager, Zhan, Koll, Bains, Petkowski, Huang, \& Lin}]{ranjanPhotochemicalRunawayExoplanet2022}
Ranjan, S., Seager, S., Zhan, Z., {et~al.} 2022, \bibinfo{title}{Photochemical {{Runaway}} in {{Exoplanet Atmospheres}}: {{Implications}} for {{Biosignatures}},} The Astrophysical Journal, 930, 131, \dodoi{10.3847/1538-4357/ac5749}

\bibitem[{T.~P. Robitaille {et~al.}(2013)Robitaille, Tollerud, Greenfield, Droettboom, Bray, Aldcroft, Davis, Ginsburg, {Price-Whelan}, Kerzendorf, Conley, Crighton, Barbary, Muna, Ferguson, Grollier, Parikh, Nair, G{\"u}nther, Deil, Woillez, Conseil, Kramer, Turner, Singer, Fox, Weaver, Zabalza, Edwards, Bostroem, Burke, Casey, Crawford, Dencheva, Ely, Jenness, Labrie, Lim, Pierfederici, Pontzen, Ptak, Refsdal, Servillat, \& Streicher}]{robitailleAstropyCommunityPython2013}
Robitaille, T.~P., Tollerud, E.~J., Greenfield, P., {et~al.} 2013, \bibinfo{title}{Astropy: {{A}} Community {{Python}} Package for Astronomy,} Astronomy \& Astrophysics, 558, A33, \dodoi{10.1051/0004-6361/201322068}

\bibitem[{U. Schumann \& H. Huntrieser(2007)Schumann \& Huntrieser}]{schumannGlobalLightninginducedNitrogen2007}
Schumann, U., \& Huntrieser, H. 2007, \bibinfo{title}{The Global Lightning-Induced Nitrogen Oxides Source,} Atmospheric Chemistry and Physics, 7, 3823, \dodoi{10.5194/acp-7-3823-2007}

\bibitem[{E.~W. Schwieterman {et~al.}(2019)Schwieterman, Reinhard, Olson, Ozaki, Harman, Hong, \& Lyons}]{schwietermanRethinkingCOAntibiosignatures2019}
Schwieterman, E.~W., Reinhard, C.~T., Olson, S.~L., {et~al.} 2019, \bibinfo{title}{Rethinking {{CO Antibiosignatures}} in the {{Search}} for {{Life Beyond}} the {{Solar System}},} The Astrophysical Journal, 874, 9, \dodoi{10.3847/1538-4357/ab05e1}

\bibitem[{E.~W. Schwieterman {et~al.}(2016)Schwieterman, Meadows, {Domagal-Goldman}, Deming, Arney, Luger, Harman, Misra, \& Barnes}]{schwietermanIdentifyingPlanetaryBiosignature2016}
Schwieterman, E.~W., Meadows, V.~S., {Domagal-Goldman}, S.~D., {et~al.} 2016, \bibinfo{title}{Identifying {{Planetary Biosignature Impostors}}: {{Spectral Features}} of {{Co}} and {{O4 Resulting From Abiotic O2}}/{{O3 Production}},} The Astrophysical Journal, 819, L13, \dodoi{10.3847/2041-8205/819/1/L13}

\bibitem[{A. Segura {et~al.}(2005)Segura, Kasting, Meadows, Cohen, Scalo, Crisp, Butler, \& Tinetti}]{seguraBiosignaturesEarthLikePlanets2005}
Segura, A., Kasting, J.~F., Meadows, V., {et~al.} 2005, \bibinfo{title}{Biosignatures from {{Earth-Like Planets Around M Dwarfs}},} Astrobiology, 5, 706, \dodoi{10.1089/ast.2005.5.706}

\bibitem[{A. Segura {et~al.}(2010)Segura, Walkowicz, Meadows, Kasting, \& Hawley}]{seguraEffectStrongStellar2010}
Segura, A., Walkowicz, L.~M., Meadows, V., Kasting, J., \& Hawley, S. 2010, \bibinfo{title}{The {{Effect}} of a {{Strong Stellar Flare}} on the {{Atmospheric Chemistry}} of an {{Earth-like Planet Orbiting}} an {{M Dwarf}},} Astrobiology, 10, 751, \dodoi{10.1089/ast.2009.0376}

\bibitem[{K. Stamnes {et~al.}(1988)Stamnes, Tsay, Jayaweera, \& Wiscombe}]{stamnesNumericallyStableAlgorithm1988}
Stamnes, K., Tsay, S.~C., Jayaweera, K., \& Wiscombe, W. 1988, \bibinfo{title}{Numerically Stable Algorithm for Discrete-Ordinate-Method Radiative Transfer in Multiple Scattering and Emitting Layered Media,} Applied Optics, 27, 2502, \dodoi{10.1364/AO.27.002502}

\bibitem[{D.~J. Teal {et~al.}(2022)Teal, Kempton, Bastelberger, Youngblood, \& Arney}]{tealEffectsUVStellar2022}
Teal, D.~J., Kempton, E. M.-R., Bastelberger, S., Youngblood, A., \& Arney, G. 2022, \bibinfo{title}{Effects of {{UV Stellar Spectral Uncertainty}} on the {{Chemistry}} of {{Terrestrial Atmospheres}},} The Astrophysical Journal, 927, 90, \dodoi{10.3847/1538-4357/ac4d99}

\bibitem[{T.~B. Thomas {et~al.}(2025)Thomas, Meadows, {Krissansen-Totton}, Gialluca, Wogan, \& Catling}]{thomasStatisticalGeochemicalConstraints2025}
Thomas, T.~B., Meadows, V.~S., {Krissansen-Totton}, J., {et~al.} 2025, \bibinfo{title}{Statistical {{Geochemical Constraints}} on {{Present-day Water Outgassing}} as a {{Source}} of {{Secondary Atmospheres}} on the {{TRAPPIST-1 Exoplanets}},} The Planetary Science Journal, 6, 126, \dodoi{10.3847/PSJ/add261}

\bibitem[{M.~A. Thompson {et~al.}(2022)Thompson, {Krissansen-Totton}, Wogan, Telus, \& Fortney}]{thompsonCaseContextAtmospheric2022}
Thompson, M.~A., {Krissansen-Totton}, J., Wogan, N., Telus, M., \& Fortney, J.~J. 2022, \bibinfo{title}{The Case and Context for Atmospheric Methane as an Exoplanet Biosignature,} Proceedings of the National Academy of Sciences, 119, e2117933119, \dodoi{10.1073/pnas.2117933119}

\bibitem[{G. Thuillier {et~al.}(2004)Thuillier, Floyd, Woods, Cebula, Hilsenrath, Hers{\'e}, \& Labs}]{thuillierSolarIrradianceReference2004}
Thuillier, G., Floyd, L., Woods, T.~N., {et~al.} 2004, \bibinfo{title}{Solar Irradiance Reference Spectra for Two Solar Active Levels,} Advances in Space Research, 34, 256, \dodoi{10.1016/j.asr.2002.12.004}

\bibitem[{F. Tian {et~al.}(2014)Tian, France, Linsky, Mauas, \& Vieytes}]{tianHighStellarFUV2014}
Tian, F., France, K., Linsky, J.~L., Mauas, P. J.~D., \& Vieytes, M.~C. 2014, \bibinfo{title}{High Stellar {{FUV}}/{{NUV}} Ratio and Oxygen Contents in the Atmospheres of Potentially Habitable Planets,} Earth and Planetary Science Letters, 385, 22, \dodoi{10.1016/j.epsl.2013.10.024}

\bibitem[{M.~G. Trainer {et~al.}(2006)Trainer, Pavlov, DeWitt, Jimenez, McKay, Toon, \& Tolbert}]{trainerOrganicHazeTitan2006}
Trainer, M.~G., Pavlov, A.~A., DeWitt, H.~L., {et~al.} 2006, \bibinfo{title}{Organic Haze on {{Titan}} and the Early {{Earth}},} Proceedings of the National Academy of Sciences, 103, 18035, \dodoi{10.1073/pnas.0608561103}

\bibitem[{G. {van Rossum} \& J. {de Boer}(1991){van Rossum} \& {de Boer}}]{vanrossumInteractivelyTestingRemote1991}
{van Rossum}, G., \& {de Boer}, J. 1991, \bibinfo{title}{Interactively Testing Remote Servers Using the {{Python}} Programming Language,} CWI Quarterly, 4, 283

\bibitem[{O. Venot {et~al.}(2016)Venot, Rocchetto, Carl, Hashim, \& Decin}]{venotINFLUENCEStelLARFLARES2016}
Venot, O., Rocchetto, M., Carl, S., Hashim, A.~R., \& Decin, L. 2016, \bibinfo{title}{{{Influence of Stellar Flares on the Chemical Composition of Exoplanets and Spectra}},} The Astrophysical Journal, 830, 77, \dodoi{10.3847/0004-637X/830/2/77}

\bibitem[{K. Vida {et~al.}(2017)Vida, K{\H o}v{\'a}ri, P{\'a}l, Ol{\'a}h, \& Kriskovics}]{vidaFrequentFlaringTRAPPIST12017}
Vida, K., K{\H o}v{\'a}ri, Z., P{\'a}l, A., Ol{\'a}h, K., \& Kriskovics, L. 2017, \bibinfo{title}{Frequent {{Flaring}} in the {{TRAPPIST-1 System}}---{{Unsuited}} for {{Life}}?} The Astrophysical Journal, 841, 124, \dodoi{10.3847/1538-4357/aa6f05}

\bibitem[{Y. Wang {et~al.}(2016)Wang, Tian, Li, \& Hu}]{wangDetectionCarbonMonoxide2016}
Wang, Y., Tian, F., Li, T., \& Hu, Y. 2016, \bibinfo{title}{On the Detection of Carbon Monoxide as an Anti-Biosignature in Exoplanetary Atmospheres,} Icarus, 266, 15, \dodoi{10.1016/j.icarus.2015.11.010}

\bibitem[{Y. Watanabe \& K. Ozaki(2024)Watanabe \& Ozaki}]{watanabeRelativeAbundancesCO22024}
Watanabe, Y., \& Ozaki, K. 2024, \bibinfo{title}{Relative {{Abundances}} of {{CO2}}, {{CO}}, and {{CH4}} in {{Atmospheres}} of {{Earth-like Lifeless Planets}},} The Astrophysical Journal, 961, 1, \dodoi{10.3847/1538-4357/ad10a2}

\bibitem[{D.~J. Wilson {et~al.}(2021)Wilson, Froning, Duvvuri, France, Youngblood, Schneider, {Berta-Thompson}, Brown, Buccino, Hawley, Irwin, Kaltenegger, Kowalski, Linsky, Loyd, Miguel, Pineda, Redfield, Roberge, Rugheimer, Tian, \& Vieytes}]{wilsonMegaMUSCLESSpectralEnergy2021}
Wilson, D.~J., Froning, C.~S., Duvvuri, G.~M., {et~al.} 2021, \bibinfo{title}{The {{Mega-MUSCLES Spectral Energy Distribution}} of {{TRAPPIST-1}},} The Astrophysical Journal, 911, 18, \dodoi{10.3847/1538-4357/abe771}

\bibitem[{N. Wogan {et~al.}(2020)Wogan, {Krissansen-Totton}, \& Catling}]{woganAbundantAtmosphericMethane2020}
Wogan, N., {Krissansen-Totton}, J., \& Catling, D.~C. 2020, \bibinfo{title}{Abundant {{Atmospheric Methane}} from {{Volcanism}} on {{Terrestrial Planets Is Unlikely}} and {{Strengthens}} the {{Case}} for {{Methane}} as a {{Biosignature}},} The Planetary Science Journal, 1, 58, \dodoi{10.3847/PSJ/abb99e}

\bibitem[{N.~F. Wogan \& D.~C. Catling(2020)Wogan \& Catling}]{woganWhenChemicalDisequilibrium2020}
Wogan, N.~F., \& Catling, D.~C. 2020, \bibinfo{title}{When Is {{Chemical Disequilibrium}} in {{Earth-like Planetary Atmospheres}} a {{Biosignature}} versus an {{Anti-biosignature}}? {{Disequilibria}} from {{Dead}} to {{Living Worlds}},} The Astrophysical Journal, 892, 127, \dodoi{10.3847/1538-4357/ab7b81}

\bibitem[{E.~T. Wolf {et~al.}(2025)Wolf, Schwieterman, {Haqq-Misra}, Fauchez, Bastelberger, Leung, Peacock, Villanueva, \& Kopparapu}]{wolfChemistryClimateTransmission2025}
Wolf, E.~T., Schwieterman, E.~W., {Haqq-Misra}, J., {et~al.} 2025, \bibinfo{title}{Chemistry, {{Climate}}, and {{Transmission Spectra}} of {{TRAPPIST-1}} e {{Explored}} with a {{Multimodel Sparse Sampled Ensemble}},} The Planetary Science Journal, 6, 231, \dodoi{10.3847/PSJ/ae031e}

\bibitem[{J.~M. Wolfe \& G.~P. Fournier(2018)Wolfe \& Fournier}]{wolfeHorizontalGeneTransfer2018}
Wolfe, J.~M., \& Fournier, G.~P. 2018, \bibinfo{title}{Horizontal Gene Transfer Constrains the Timing of Methanogen Evolution,} Nature Ecology \& Evolution, 2, 897, \dodoi{10.1038/s41559-018-0513-7}

\bibitem[{M.~L. Wong {et~al.}(2025)Wong, Prabhu, Alexander, Cleaves, Cody, Hystad, Bermanec, Bleeker, Boyce, Corpolongo, Czaja, Das, Gaines, Gregory, Jaszczak, Javaux, Jodder, Knoll, Van~Kranendonk, Maloney, Noffke, Rainbird, Slaughter, St{\"u}eken, Summons, Westall, Wiemann, Xiao, \& Hazen}]{wongOrganicGeochemicalEvidence2025}
Wong, M.~L., Prabhu, A., Alexander, C.~O., {et~al.} 2025, \bibinfo{title}{Organic Geochemical Evidence for Life in {{Archean}} Rocks Identified by Pyrolysis--{{GC}}--{{MS}} and Supervised Machine Learning,} Proceedings of the National Academy of Sciences, 122, e2514534122, \dodoi{10.1073/pnas.2514534122}

\bibitem[{R. Wordsworth(2015)Wordsworth}]{wordsworthATMOSPHERICHEATREDISTRIBUTION2015}
Wordsworth, R. 2015, \bibinfo{title}{{{Atmospheric Heat Redistribution and Collapse on Tidally Locked Rocky Planets}},} The Astrophysical Journal, 806, 180, \dodoi{10.1088/0004-637X/806/2/180}

\bibitem[{F. Wunderlich {et~al.}(2020)Wunderlich, Scheucher, Godolt, Grenfell, Schreier, Schneider, Wilson, {S{\'a}nchez-L{\'o}pez}, {L{\'o}pez-Puertas}, \& Rauer}]{wunderlichDistinguishingWetDry2020}
Wunderlich, F., Scheucher, M., Godolt, M., {et~al.} 2020, \bibinfo{title}{Distinguishing between {{Wet}} and {{Dry Atmospheres}} of {{TRAPPIST-1}} e and f,} The Astrophysical Journal, 901, 126, \dodoi{10.3847/1538-4357/aba59c}

\bibitem[{K. Zahnle {et~al.}(2008)Zahnle, Haberle, Catling, \& Kasting}]{zahnlePhotochemicalInstabilityAncient2008}
Zahnle, K., Haberle, R.~M., Catling, D.~C., \& Kasting, J.~F. 2008, \bibinfo{title}{Photochemical Instability of the Ancient {{Martian}} Atmosphere,} Journal of Geophysical Research: Planets, 113, \dodoi{10.1029/2008JE003160}

\bibitem[{K.~J. Zahnle \& D.~C. Catling(2017)Zahnle \& Catling}]{zahnleCosmicShorelineEvidence2017}
Zahnle, K.~J., \& Catling, D.~C. 2017, \bibinfo{title}{The {{Cosmic Shoreline}}: {{The Evidence}} That {{Escape Determines}} Which {{Planets Have Atmospheres}}, and What This {{May Mean}} for {{Proxima Centauri B}},} The Astrophysical Journal, 843, 122, \dodoi{10.3847/1538-4357/aa7846}

\bibitem[{K.~J. Zahnle {et~al.}(2006)Zahnle, Claire, \& Catling}]{zahnleLossMassindependentFractionation2006}
Zahnle, K.~J., Claire, M., \& Catling, D. 2006, \bibinfo{title}{The Loss of Mass-Independent Fractionation in Sulfur Due to a {{Palaeoproterozoic}} Collapse of Atmospheric Methane,} Geobiology, 4, 271, \dodoi{10.1111/j.1472-4669.2006.00085.x}

\bibitem[{Z. Zhan {et~al.}(2022)Zhan, Huang, Seager, Petkowski, \& Ranjan}]{zhanOrganicCarbonylsAre2022}
Zhan, Z., Huang, J., Seager, S., Petkowski, J.~J., \& Ranjan, S. 2022, \bibinfo{title}{Organic {{Carbonyls Are Poor Biosignature Gases}} in {{Exoplanet Atmospheres}} but {{May Generate Significant CO}},} The Astrophysical Journal, 930, 133, \dodoi{10.3847/1538-4357/ac64a8}

\bibitem[{S. Zieba {et~al.}(2023)Zieba, Kreidberg, Ducrot, Gillon, Morley, Schaefer, Tamburo, Koll, Lyu, Acu{\~n}a, Agol, Iyer, Hu, Lincowski, Meadows, Selsis, Bolmont, Mandell, \& Suissa}]{ziebaNoThickCarbon2023}
Zieba, S., Kreidberg, L., Ducrot, E., {et~al.} 2023, \bibinfo{title}{No Thick Carbon Dioxide Atmosphere on the Rocky Exoplanet {{TRAPPIST-1}} c,} Nature, 620, 746, \dodoi{10.1038/s41586-023-06232-z}

\end{thebibliography}
\bibliographystyle{aasjournalv7}


\end{document}